\documentclass[twocolumn]{aastex62}

\newcommand{\Msun}{M_{\odot}}

\newcommand{\Mstar}{M_{\star}}
\newcommand{\Mh}{M_{\rm halo}}

\newcommand{\hinv}{h^{-1}}
\newcommand{\himpc}{{h^{-1}\,{\rm Mpc}}}
\newcommand{\hikpc}{{h^{-1}\,{\rm kpc}}}
\newcommand{\hicmpc}{{h^{-1}\,{\rm cMpc}}}
\newcommand{\hickpc}{{h^{-1}\,{\rm ckpc}}}
\newcommand{\himsun}{{h^{-1}\,{\Msun}}}
\newcommand{\HI}{{\sc Hi}}
\newcommand{\NHI}{N_{\rm HI}}
\newcommand{\Lya}{{{\rm Ly}\alpha}}
\newcommand{\delgal}{\delta_{\rm gal}}
\newcommand{\Pk}{P_{\rm 1D}(k)}
\newcommand{\kms}{{\rm km\,s^{-1}}}
\graphicspath{{./}{figures/}}

\shorttitle{Probing Feedback via IGM tomography}
\shortauthors{Nagamine et al.}
\begin{document}

\title{Probing Feedback via IGM tomography and the Ly$\alpha$ Forest with Subaru PFS, TMT/ELT, and JWST}

\correspondingauthor{Kentaro Nagamine}
\email{kn@astro-osaka.jp}

\author[0000-0001-7457-8487 ]{Kentaro Nagamine}
\affil{Theoretical Astrophysics, Department of Earth and Space Science, Graduate School of Science, Osaka University, \\
1-1 Machikaneyama, Toyonaka, Osaka 560-0043, Japan}
\affil{Kavli-IPMU (WPI), University of Tokyo, 5-1-5 Kashiwanoha, Kashiwa, Chiba, 277-8583, Japan}
\affil{Department of Physics \& Astronomy, University of Nevada, Las Vegas, 4505 S. Maryland Pkwy, Las Vegas, NV 89154-4002, USA}

\author{Ikkoh Shimizu}
\affil{Shikoku Gakuin University, 3-2-1 Bunkyocho, Zentsuji, Kagawa, 765-8505, Japan}
%\collaboration{(AAS Journals Data Scientists collaboration)}

\author{Katsumi Fujita}
\affiliation{Theoretical Astrophysics, Department of Earth and Space Science, Graduate School of Science, Osaka University, \\
1-1 Machikaneyama, Toyonaka, Osaka 560-0043, Japan}
%\nocollaboration

\author{Nao Suzuki}
\affil{Kavli-IPMU (WPI), University of Tokyo, 5-1-5 Kashiwanoha, Kashiwa, Chiba, 277-8583, Japan}
%\collaboration{(LaTeX collaboration)}

\author{Khee-Gan Lee}
\affil{Kavli-IPMU (WPI), University of Tokyo, 5-1-5 Kashiwanoha, Kashiwa, Chiba, 277-8583, Japan}

\author{Rieko Momose}
\affil{Department of Astronomy, University of Tokyo, 7-3-1 Hongo, Bunkyo, Tokyo, 113-8654, Japan}

\author{Shiro Mukae}
\affil{ICRR, University of Tokyo, 5-1-5 Kashiwanoha, Kashiwa, Chiba, 277-8582, Japan}
\affil{Department of Astronomy, University of Tokyo, 7-3-1 Hongo, Bunkyo, Tokyo, 113-8654, Japan}

\author{Yongming Liang}
\affil{Department of Astronomy, University of Tokyo, 7-3-1 Hongo, Bunkyo, Tokyo, 113-8654, Japan}
\affil{National Astronomical Observatory of Japan, Mitaka, Tokyo, Japan}

\author{Nobunari Kashikawa}
\affil{Department of Astronomy, University of Tokyo, 7-3-1 Hongo, Bunkyo, Tokyo, 113-8654, Japan}

\author{Masami Ouchi}
\affil{National Astronomical Observatory of Japan, Mitaka, Tokyo, Japan}
\affil{Department of Astronomy, University of Tokyo, 7-3-1 Hongo, Bunkyo, Tokyo, 113-8654, Japan}\affil{ICRR, University of Tokyo, 5-1-5 Kashiwanoha, Kashiwa, Chiba, 277-8582, Japan}

\author{John D. Silverman}
\affil{Kavli-IPMU (WPI), University of Tokyo, 5-1-5 Kashiwanoha, Kashiwa, Chiba, 277-8583, Japan}



\begin{abstract}

In preparation for the IGM tomography study by Subaru Prime Focus Spectrograph (PFS) survey and other large future telescopes such as TMT/ELT/GMT, we present the results of our pilot study on $\Lya$ forest and IGM tomography statistics using the {\sc GADGET3-Osaka}  cosmological smoothed particle hydrodynamical simulation.  
Our simulation includes models for star formation and supernova feedback, which enables more realistic cross-correlation studies between galaxies, neutral hydrogen ({\HI}) and metals in circumgalactic and intergalactic medium. We create a light-cone data set at $z=2-3$ from our simulations and generate mock $\Lya$ forest data.
As a first step, in this paper, we focus on the distribution of {\HI} and galaxies, and present statistical results on 1D flux PDF, 1D power spectrum, flux contrast vs. impact parameter, {\HI}--galaxy cross-correlations. Our results show overall agreement with current observational data, with some interesting discrepancies on small scales that are due to either feedback effects or varying observational conditions. 
Our simulation shows stronger {\HI} absorption with decreasing transverse distance from galaxies.   
We find that the massive galaxies with $\Mstar \ge 10^{10}\,\Msun$ contribute strongly to the flux contrast signal, and that the lower-mass galaxies with $\Mstar \sim 10^8-10^{10}\,\Msun$ tend to dilute the flux contrast signal from massive galaxies. 
On large scales, the average flux contrast smoothly connects to the IGM level, supporting the concordance $\Lambda$ cold dark matter model. 
We also find an increase in the {\HI} absorption toward the center of a protocluster. 

%AAS Journals, the Astrophysical Journal (ApJ), the Astrophysical Journal Letters (ApJL), and Astronomical Journal (AJ), all have a 250 word limit for the abstract

\end{abstract}


\section{Introduction} \label{sec:intro}

Understanding the distribution of baryons and galaxies is one of the most important topics in modern cosmological studies.  In particular, hydrogen is the most abundant element in our universe, and the distribution of neutral hydrogen (\HI) contains useful information on the radiative history of our universe, e.g., evolution of the ultraviolet background (UVB) radiation field, the cosmic star formation history, and the interaction between galaxies, circum-galactic medium \citep[CGM; e.g.,][]{Tumlinson17} and inter-galactic medium \citep[IGM; e.g.,][]{Meiksin09,McQuinn16}.

\HI\ gas at cosmological distances can be observed as the Lyman-$\alpha$ (hereafter $\Lya$) forest in quasar absorption lines, and abundant data from high-resolution spectroscopy have been collected over the past few decades \citep[e.g.,][]{Weymann81,Cowie95,Rauch98}. In conjunction with those observational efforts, cosmological hydrodynamic simulations have played crucial roles in deepening our understanding of the nature of the $\Lya$ forest clouds \citep{Cen94,Hernquist96,Miralda96,Zhang97,Zhang98},  
and it is generally accepted that the $\Lya$ forest originates from the diffuse \HI\ gas in filamentary structures as well as the gaseous clouds mildly bound by gravity (Ly-limit systems). 
The $\Lya$ forest has proven to be one of the most powerful probes of cosmology, and it has been used to constrain the cosmological parameters \citep{Weinberg98a,McDonald06}, the matter power spectrum \citep{Croft98,Irsic17}, the mass of warm dark matter particles \citep[e.g.][]{Viel05,Viel13a}, the mass of neutrinos \citep{Palanque15b}, and the impact of supernova feedback on IGM \citep[e.g.][]{Theuns02,Cen05,Kollmeier06}. 

Over the past several years, by utilizing a large quasar catalog from the Sloan Digital Sky Survey--Baryon Oscillation Spectroscopic Survey \citep[SDSS-BOSS;][$\sim 160,000$ sightlines]{Dawson13}, significant {\sc Hi} overdensities (i.e. protocluster candidates) have been identified at $z=2-3$ \citep{Cai16,Cai17,Mukae20,Ravoux20}, which are expected to be a more uniform, unbiased sample of protoclusters, covering a large cosmological volume of $\sim$\,1\,Gpc$^3$.  
(However, note the cautionary remarks by \citealt{Miller19} as well. See also Section\,\ref{sec:proto} for related discussions.)
These protoclusters will serve as unique test beds for the hierarchical structure formation scenario \citep{Overzier16,Chiang17}.  
For example, the most massive galaxies are found in protoclusters, and they are expected to be the first ones to make the transition from the star-forming blue sequence to the red sequence.  The connections between radio galaxies, $\Lya$ blobs, and protoclusters are also of significant interest \citep[e.g.,][]{Umehata15,Umehata20}. 

One can also perform the ``IGM tomography" using more ubiquitous bright star-forming galaxies (e.g., Ly-break galaxies (LBGs) at $z\sim 3$) as background sources, and measure the 3D distribution of \HI\ gas in the foreground at $z=2-3$. Indeed, \citet{Lee14b} have demonstrated that this can be done using the Keck telescope, i.e., the CLAMATO survey \citep{Lee18}. 
A galaxy protocluster has been identified at $z=2.44$ by the $\Lya$ tomographic technique \citep{Lee16a}. 
The scientific goals of IGM tomography are: (i) to characterize the cosmic web at $z>2$, (ii) to study the association between galaxies/active galactic nuclei (AGNs) and \HI\ gas, and (iii) to identify protoclusters and voids in an unbiased fashion. 

There are many large spectroscopic surveys being planned in the next decade to probe the intermediate redshift range of $z>1$.  The SDSS and 2-degree Field (2dF) surveys gave us excellent views of our local universe, and the ``SDSS at $z\sim  1$" will soon become available by the combination of Subaru Hyper-Supreme Cam (HSC) and Prime Focus Spectrograph (PFS) projects \citep{Takada14}.  
The Subaru PFS survey is a large spectroscopic survey on the Subaru telescope employing $\sim 2400$ fibers positioned across a 1.3\,deg field, which is scheduled to start from $\sim$\,2022. 
It follows up on the ongoing HSC imaging survey, and the combination of the two large projects will present unique opportunities to study galaxy evolution and {\HI} distribution in our universe at the same time. 
The high-$z$ program of the PFS project plans to cover the sky area of 15 deg$^2$ of the HSC deep fields, and is likely to include topics such as the IGM tomography at $z=2-3$, galaxy evolution at $z=2-6$, reionization studies using $\Lya$ emitters (LAEs) at $z=6-7$.

While there are other similar projects to the Subaru PFS utilizing a multiplexed fiber spectrograph, such as the WEAVE \citep{Dalton12} and the MOONS \citep{Cirasuolo14}, PFS has a unique combination of both telescope diameter and spectral coverage into the blue. 
This allows it to observe large numbers of galaxies at $z\sim 2-3$ and at the same time measure the Ly$\alpha$ forest absorption with large numbers of background sources. 

In this paper, we focus on the science cases related to the PFS IGM tomography program, and prepare to make some forecasts using cosmological hydrodynamic simulations that include full physics of star formation and supernova (SN) feedback. 
Some details of the Subaru PFS project are summarized in Appendix~\ref{app:obs}. 

For the IGM tomography, the spatial resolution of current observational studies is still coarse, and numerical simulations can provide useful comparison data set, by mimicking the actual observations to evaluate the expected observational results. 
For example, \citet{Lukic15} used the {\small NYX} hydrodynamic simulation to examine $\Lya$ forest statistics,  but without generating a full light-cone dataset nor the treatment of star formation and feedback (i.e., an optically thin calculation).  They found that a hydrodynamic resolution of $20\,\hikpc$ is required to achieve 1\% convergence of $\Lya$ forest flux statistics up to $k=10\,h$\,Mpc$^{-1}$, 
and box sizes of $L>40\,\hicmpc$ (cMpc denotes comoving Mpc) are needed to suppress the errors below 1\% for the 1D flux power spectrum. Typically, convergence of $5\%-10\%$ had been reported by several authors using box sizes of $20-40\,\hicmpc$ with resolutions of $50-200\,\hickpc$ at $z\sim 2$ \citep[e.g.,][]{Meiksin04,McDonald05b,Viel06,Bolton09,Tytler09,Meiksin14}. 

Furthermore, \citet{Sorini18} used the Illustris and {\small NYX} simulations to examine the average $\Lya$ absorption profile around galaxies, and showed that the results agree well with the observations by BOSS \citep{Font-Ribera13} and quasar pairs \citep{Pro13,Rubin15} at transverse distances of 2\,Mpc\,$\lesssim b \lesssim 20$\,Mpc. They argued that the nice asymptotic agreement of the absorption profile on large scales is a reflection of the fact that the $\Lambda$ cold dark matter (CDM) model successfully describes the distribution of ambient IGM around the dark matter halos. Based on the comparison between the simulations and observations, they also suggested that the `sphere of influence' of galaxies could extend out to $\sim$\,7 times the halo virial radius, i.e. to $\sim$\,2\,Mpc, and that there are significant differences between the simulations and observations on small scales of $\lesssim$\,100\,kpc. 
Earlier similar works using hydrodynamic simulations include e.g., \citet[][]{Bruscoli03,Kollmeier03,Kollmeier06,Meiksin15,Meiksin17}.
\citet{Sorini18} also showed that the both Illustris and {\small NYX} simulations underpredict the $\Lya$ absorption profile around quasars and LBGs at small separations of $b<R_{\rm vir}$. 
\citet{Sorini20} further examined the {\small SIMBA} hydrodynamic simulation 
%\citep{Dave19} 
and showed that it overpredicts the absorption (i.e. flux contrast) on small scales of $b<100$\,kpc significantly.
We show in this paper that our {\sc GADGET3-Osaka} simulation does not have these problems on small scales, and we highlight the variation caused by the differences in feedback models and galaxy samples. 

An important difference of the present work from earlier ones that adopted the  optically thin fluctuating Gunn--Peterson approximation (FGPA) treatment \citep[][]{Croft98,Weinberg98a} or from those without detailed treatment of star formation and feedback, is that we use cosmological hydrodynamic simulations with detailed models of star formation and SN feedback (but no AGN feedback yet). 
For example, the {\sc Sherwood} simulations \citep{Bolton17} operate at slightly higher resolution than our simulation, but they did not treat the star formation and metal enrichment process in detail, and instead resorted to a simplified model of converting the densest gas into star particles to save computing time \citep[see also][]{Meiksin14,Meiksin15,Meiksin17}. 

We note that \citet{Sorini20} found that the impact of AGN feedback is not so strong, and that the stellar feedback is the primary driver for determining the average physical properties of CGM at $z=2-3$.  Their claim needs to be checked in other simulations with AGN feedback in the future, but the absence of AGN feedback treatment in the present paper is perhaps not so critical, because we are not discussing the proximity effect around quasars in particular (but see the discussion in Section\,\ref{sec:proto}). 

As a pathfinder to the IGM tomography studies by the Subaru PFS and upcoming large telescopes such as TMT/ELT, we examine various statistics related to the $\Lya$ forest, such as the 1D flux probability distribution function (PDF), 1D power spectrum, $\Lya$ decrement as a function of transverse distance from galaxies, and cross-correlation between galaxies and {\sc Hi} gas. This is only our first step toward more rigorous comparisons between simulations and observations of CGM and IGM. 
In this paper, we focus on the relative distribution of galaxies and {\HI}, and do not touch on the metal absorption lines, which we will discuss in our later publications. 
The goal of this paper is not to argue for a best-fit simulation parameter set  or subgrid models, but rather to highlight the differences from the currently available observational data and look for the right directions for our future research. 
The astrophysics in this subject is quite rich, as we need to have a good understanding of cosmology, galaxy formation, star formation, feedback, and chemical enrichment of the IGM, and link them all together to have a full picture. 

The results presented in this paper serve as the basis for our future work of comparing the simulations and observations, for example, as presented by \citet{Momose21a,Momose21b,Liang21}.  We will refer to these results in more detail in later sections. 

This paper is organized as follows. We describe the details of our cosmological hydrodynamic simulations in Section~\ref{sec:sim}, and the method of generating the Ly$\alpha$ forest data is presented in Section~\ref{sec:method}. 
In Section~\ref{sec:result}, we lay out the results of $\Lya$ flux PDF, 1D $\Lya$ power spectrum, flux contrast vs. impact parameter from galaxies, cross-correlation between galaxies and {\sc Hi},  correlation between galaxy overdensity and $\Lya$ absorption decrement, and finally the flux contrast in a protocluster. 
We then present our discussion and summary in Section~\ref{sec:conclusion}.

%%%%%%%%%%%%%%%%%%%%%%%%%%%%%%%%%%%%%%%%%%%%%%%%%%

\section{Simulations and Methods}

\subsection{Cosmological Hydrodynamic Simulations}  
\label{sec:sim} 

We use the {\sc GADGET3-Osaka} cosmological smoothed particle hydrodynamics (SPH) code \citep{Aoyama17,Shimizu19}, which is a modified version of {\small GADGET-3} \citep[originally described in][as {\small GADGET-2}]{Springel05e}.  
Our code includes important improvements such as the density-independent, pressure--entropy formulation of SPH \citep[][]{Hopk13b,Saitoh13}, the time-step limiter \citep{Saitoh09b}, and a quintic spline kernel \citep{Morris96}. 

Our code also includes models for star formation and SN feedback. 
We adopt the basic model of star formation in the AGORA project \citep{Kim14,Kim16}. 
The SN feedback recipe is described in detail by \citet{Shimizu19}, which we briefly summarize here. 
The feedback energy and metals are deposited within a hot bubble radius computed from a Sedov--Taylor-like solution that takes the cooling into account \citep{Chevalier74,McKee77}, similarly to the blast-wave model by \citet{Stinson06}. 
We also use the {\small CELib} chemical evolution library by \citet{Saitoh17} and appropriate time delays for Type Ia \& II SNe, and asymptotic giant branch (AGB) stars. 
The energy and metals are injected into the surrounding ISM with appropriate time delays for different sources based on CELib, following each star formation event. 
We compute the total feedback energy for each star-forming event, and deposit both thermal (70\%) and kinetic (30\%) energy to the neighboring gas particle within the hot bubble radius, as in the fiducial model of \citet{Shimizu19}. Through the tests with isolated AGORA galaxies, we have found that this fiducial model has a moderate degree of chemical enrichment without overheating the IGM, unlike the original model of constant wind velocity by \citet{Springel03b}. The merit of the Osaka feedback model is that the wind velocity and the hot bubble size are determined based on the local physical quantities (gas density, pressure, and available feedback energy) rather than the bulk quantities such as galaxy stellar mass or halo mass, and this is more favorable for future higher-resolution simulations. 

The uniform UV radiation background model of \citet{Haardt12} is adopted. 
In one of the simulation runs, the self-shielding by optically thick gas is treated following the prescription of \citet[][which is based on the earlier similar works by \citealt{Nag10b,Altay11a,Yajima12a,Bird13}]{Rahmati13} to test its impact on the $\Lya$ forest statistics. 
The cooling is solved by the Grackle chemistry and cooling library \citep{Smith17}\footnote{https://grackle.readthedocs.org/}
with the option of ``\texttt{primordial\_chemistry\,=\,3}," which computes the detailed non-equilibrium chemistry network of 12 species. Metal line cooling is also solved. 

\begin{deluxetable}{lc}[t]
\tablecolumns{2}
%\tablenum{2}
\tablewidth{0pt}
\tablecaption{List of Numerical Simulations \label{tab:sim}}
\tablehead{
\colhead{Model} & \colhead{Notes} \\
}
%\colnumbers
\startdata
		Osaka20-Fiducial & No self-shielding \\
		Osaka20-Shield & With self-shielding \\
		Osaka20-NoFB & No SN feedback \\
		Osaka20-CW & Constant-velocity galactic wind model\tablenotemark{a} \\ %of \citet{Springel03b} 
		Osaka20-FG09 & UVB model of FG09\tablenotemark{b} \\ %\citet{Faucher09}  
\enddata
\tablenotetext{a}{\citet[][SH03]{Springel03b}}
\tablenotetext{b}{\citet[][FG09]{Faucher09}}
\tablecomments{Model description of the Osaka20 simulation series. All simulations use a box size of comoving $100\,\himpc$ (cMpc), and $2\times 512^3$ particles for gas and dark matter. %Particle masses are $m_{\rm DM} = 5.38\times 10^8\,\himsun$ and $m_{\rm gas}=1.00\times 10^8\himsun$. 
Further details of simulation parameters are given in Table~\ref{tab:res} of Appendix~\ref{app:res}. See the main text for the details of baryonic spatial resolution. 
The first four runs adopt the \citet{Haardt12} UVB model. 
See Appendix~\ref{app:res} for the discussion of the box-size effect and resolution test.
}
\end{deluxetable}

In this paper, we primarily use cosmological hydrodynamic simulations with a box size of 100\,$\hicmpc$ with a total initial particle number of $2\times 512^3$. 
We also use another simulation with a box size of 
50\,$\hicmpc$ and $2\times 256^3$ particles to examine the box-size effect, but the results are very similar and we do not show the comparison here. 
Given the above parameters, the two simulations have exactly the same mass/spatial resolution; however, the 50\,$\hicmpc$ box is slightly too small to create the light-cone data set for the $\Lya$ forest study at $z\approx 2-3$. 
We find that we have sufficient resolution to resolve the $\Lya$ forest in the wavelength space compared to the observations as we demonstrate later. 
The initial particle masses in the two simulations are $1.00\times 10^8\,\himsun$ and $5.38\times 10^8\,\himsun$ for gas and dark matter particles.  
The gas particle mass can change due to star formation and feedback. 
The gravitational softening length is set to $\epsilon_g = 7.8\,\hickpc$, but we allow the baryonic smoothing length %$h_s$ 
to become as small as $0.1 \epsilon_g$.  
This means that the minimum baryonic smoothing at $z=2$ is about physical $260\,\hinv$\,pc, which is sufficient to resolve the structures associated with the $\Lya$ forest.
We adopt the following cosmological parameters from \citet{Planck16cosmo}: ($\Omega_{\rm m}, \Omega_{\rm dm}, \Omega_{\rm b}, \sigma_8, h) = (0.3089, 0.2603, 0.04860, 0.8150, 0.6776).$

The list of simulations with different models is summarized in Table~\ref{tab:sim}.
Since the self-shielding of UVB might affect the $\Lya$ forest via star formation and feedback  \citep[e.g.,][]{Rahmati15}, we perform one run with and one without self-shielding: ``Osaka20-Shield" and ``Osaka20-Fiducial," respectively. 
The third run (Osaka20-NoFB) in Table~\ref{tab:sim} is without SN feedback to test the impact of the Osaka feedback model on the $\Lya$ forest statistics.  The fourth run (Osaka20-CW) is with the constant-velocity galactic wind model by \citet{Springel03b}. 
The fifth run (Osaka20-FG09) uses the UVB model of \citet{Faucher09} instead of that of \citet{Haardt12} to examine the impact of slight differences in the UVB model. 

% Fig. 1
\begin{figure*}
\epsscale{1.2}
\plotone{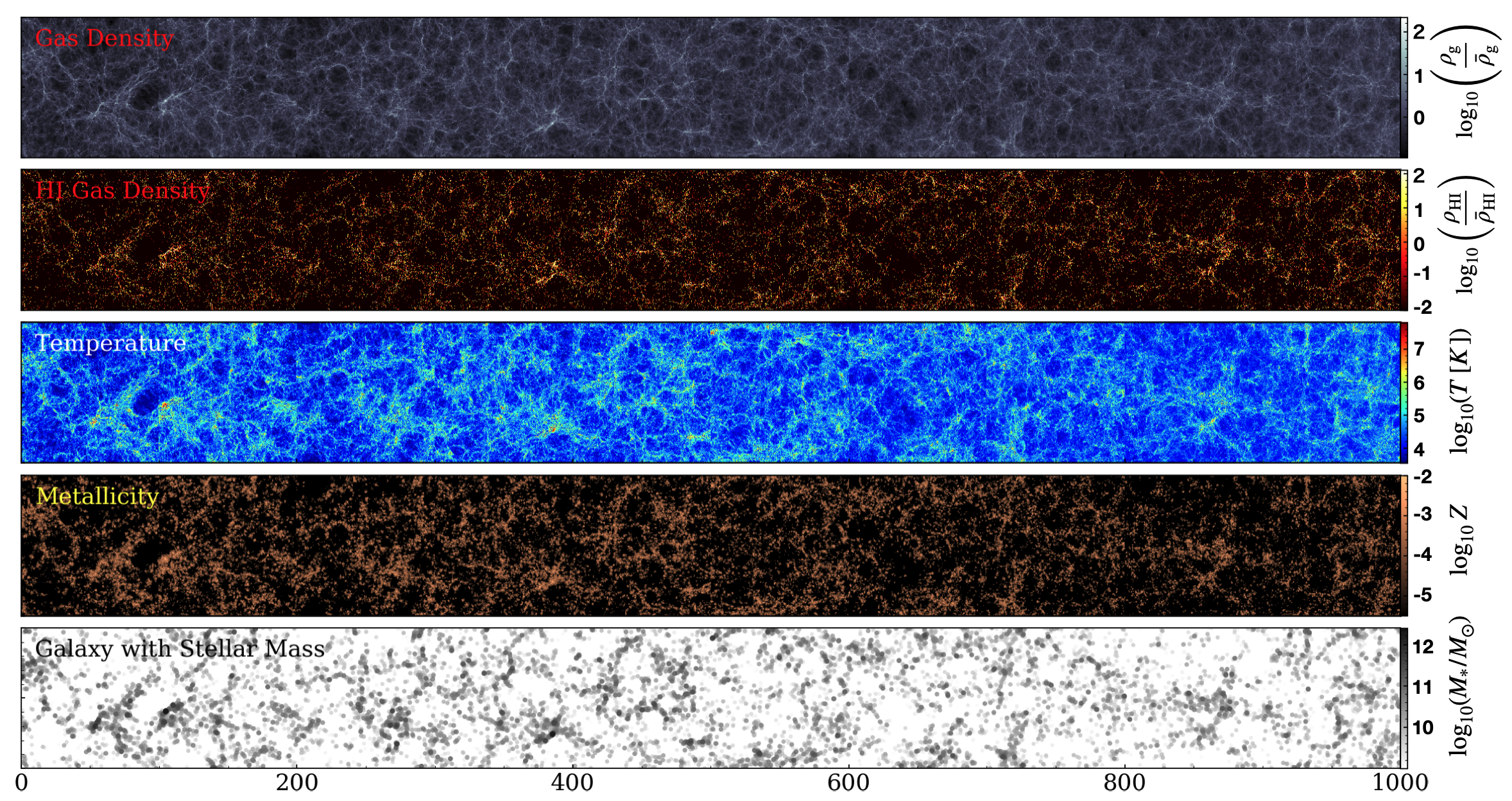}
  \caption{Light-cone output at $z\approx 2-3$, covering $100\,\hicmpc\, {\rm (height)} \times 1\,h^{-1}{\rm cGpc} \times 10\,\hicmpc$ (depth).
  Panels from top to bottom: projected gas overdensity, {\sc Hi} overdensity, temperature, metallicity, and galaxy distribution color-coded by the stellar mass, respectively.
  In the bottom panel, one can see the effect of galaxy bias, where more massive galaxies are more clustered in high-density regions. 
  }
   \label{fig:Lightcone}
\end{figure*}

% Fig. 2
\begin{figure*}
\epsscale{1.1}
\plotone{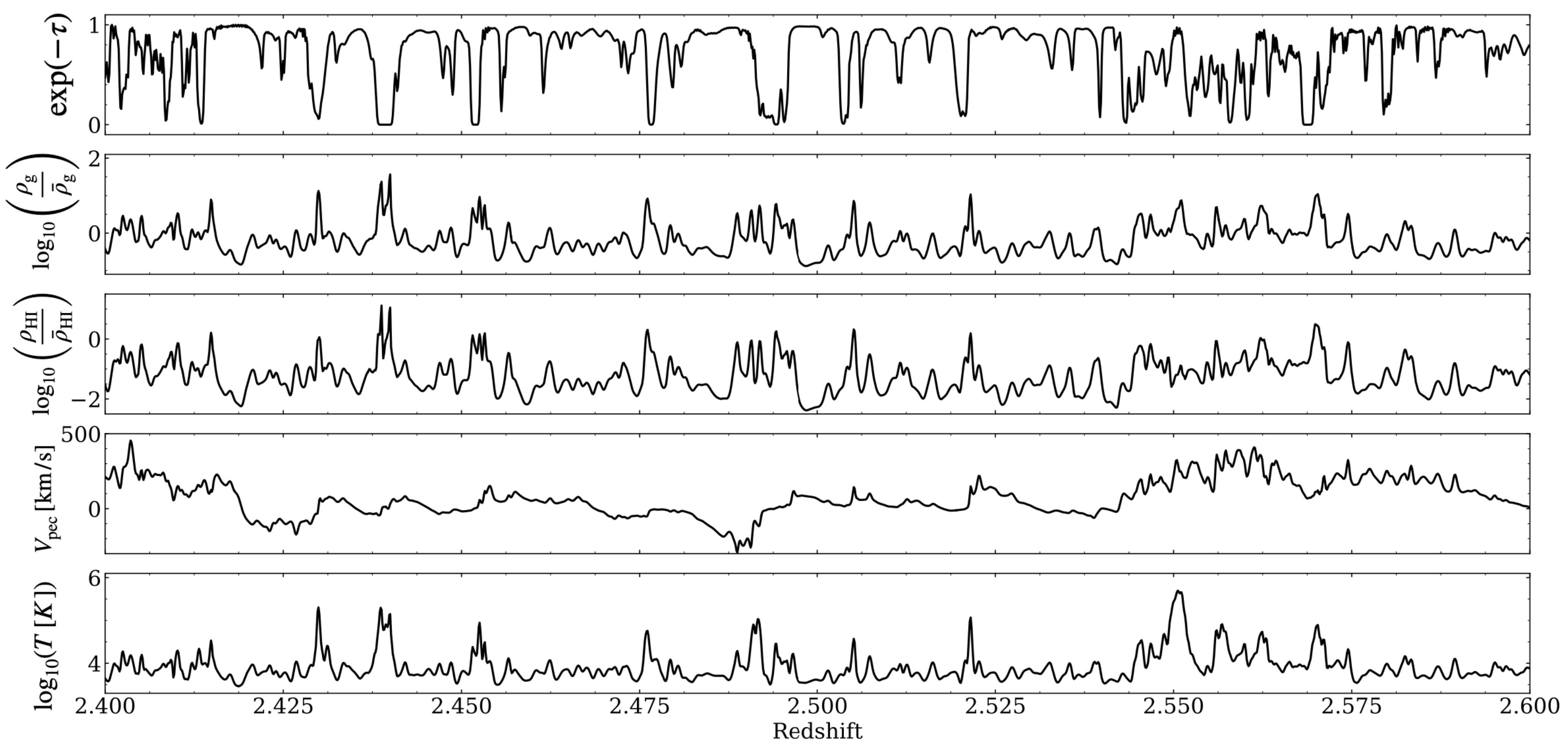}
  \caption{An example of line-of-sight data for $z=2.4-2.6$. From top to bottom: transmitted flux (i.e. $\exp(-\tau)$), total gas overdensity, {\sc Hi} overdensity, peculiar velocity along the LOS, and temperature distribution. One can see that the absorption trough is slightly shifted from the overdensity peaks due to peculiar velocities, and the \HI\ overdensities sometimes do not match with the total gas overdensities in hot overdense regions (e.g. see the features associated with a high-$T$ peak at $z\sim 2.55$.) }
    \label{fig:LOS}
\end{figure*}

%%%%%%%%%%%%%%%%%%%%%%%%%%%%%%%%%%

\subsection{Light-cone and Line-of-sight Data Set}
\label{sec:method}

Using the output of our cosmological SPH simulations, we first produce the light-cone data set for a redshift path of $z\approx 2-3$ which is the primary redshift range of the observed $\Lya$ forest, because it conveniently falls into the optical wavelength range for  ground-based spectroscopic observations.  
This is also the target redshift range for the IGM tomography by PFS \citep{Takada14}.  With the James Webb Space Telescope (JWST), we will be able to probe a higher redshift range as well, but for the moment we focus on $z=2-3$ in the present paper. 
We create the light-cone data by connecting $10$ simulation boxes of different redshifts following \citet{Shimizu14}, and cover from $z \sim 1.8$ to $3.1$.  
When connecting the simulation boxes, we randomly shift and rotate each box so that the same structures do not repeat on a single line of sight (LOS). 
This method causes discontinuities between the connected simulation boxes and it is a concern; however, we have checked that our statistical results such as the $\Lya$ power spectrum do not change when we change the details of how we connect the boxes.  In the future, we will compare the results with other methods such as the one used by \citet{Dave10} and examine the impact on the $\Lya$ forest more carefully. 
The resulting light-cone has transverse dimensions of $100\,\hicmpc \times 100\,\hicmpc$ and a $1\,h^{-1}{\rm cGpc}$ sightline. 

We then calculate the Ly$\alpha$ optical depth ($\tau$) along the LOS. 
First, we calculate the physical quantities, $A_{\rm pixel}(x)$, at each pixel $x$ along the LOS, such as {\HI} density, LOS velocity, and temperature, as follows: 
\begin{equation}
A_{\rm pixel}(x) = \sum_j \frac{m_j}{\rho_j} A_j W(r, h_j), 
\end{equation}
where $A_j$, $m_j$, $\rho_j$ and $h_j$ are the physical quantity of concern, gas particle mass, gas density, and smoothing length of the $j$-th particle, respectively. 
$W$ is the SPH kernel function, and $r$ is the distance between LOS pixel and gas particles. 
The pixel length ($dl$) is set to a constant value of $100\,h^{-1}$\,ckpc, which is a higher resolution than any of the relevant Ly$\alpha$ observations. 
Then, we calculate the Ly$\alpha$ optical depth $\tau(x)$ using these physical values at each pixel as 
\begin{equation}
\tau(x) = \frac{\pi e^2}{m_e c} \sum_j f\, \phi(x - x_j) n_{\rm HI}(x_j) dl,  
\end{equation}
where $e$, $m_e$, $c$, $f$, $n_{\rm HI}$, and $x_j$ are the electron charge, electron mass, speed of light, absorption oscillator strength, {\HI} number density, and $j$-th pixel location, respectively. 
$\phi$ is the Voigt profile, and we use the fitting formula of \citet{Tasitsiomi06b} without direct integration. 
We draw 1024 $(=32^2)$ LOSs with regularly spaced intervals, resulting in mean transverse separation of 3.3\,$h^{-1}$\,Mpc which is comparable to the CLAMATO survey. % \citep{Lee14b}.

Figure~\ref{fig:Lightcone} shows various physical quantities in the entire light-cone at $z=2-3$, covering $100\,\himpc$ (vertical) $\times$ $1\,\hinv$\,cGpc (horizontal path length) $\times$  $10\hinv$\,cMpc (depth).  
From top to bottom, the panels show total gas overdensity, \HI\ overdensity, temperature, metallicity, and the galaxy distribution color-coded by galaxy stellar masses.   
One can see visually that the higher-mass galaxies are more clustered in higher-density regions as expected from their greater clustering strength.  
Since our simulation box is limited to $100\,\hicmpc$, we do not have very massive galaxies with stellar masses $M_\star > 10^{12}\,\Msun$ at $z\sim 2$.  Ideally we would like to simulate larger volumes in the future; however, the box size of $100-200\,\hicmpc$ is currently the sweet spot due to the balance of numerical resolution, large-scale modes of the power spectrum, and the computing power of current supercomputers.

Figure~\ref{fig:LOS} shows an example of LOS data as a function of redshift at $z=2.4-2.6$, giving a more detailed view of absorption line profiles and density fluctuations.  
The five panels show, from top to bottom, the normalized transmitted flux (i.e., $\exp(-\tau)$), total gas overdensity, \HI\  overdensity, baryon peculiar velocities, and temperature profiles. 
One can see that some absorption troughs are slightly shifted from the density peaks due to peculiar velocities.  When the peculiar velocity is positive (negative), the absorption is redshifted (blueshifted). 

Just to walk the reader through some numbers, in the flat Planck cosmology adopted in this paper, the redshift path of $\Delta z \sim 0.1$ corresponds to a comoving distance of $\Delta r \approx c \Delta z / H(z) \approx c \Delta z /(3.58 H_0) \approx 120$\,cMpc at $z=2.4$, which roughly corresponds to one simulation box size. Therefore Figure~\ref{fig:LOS} is created by using about two simulation boxes of $100\,\hicmpc$ size.

%Fig. 3
\begin{figure*}
\gridline{\fig{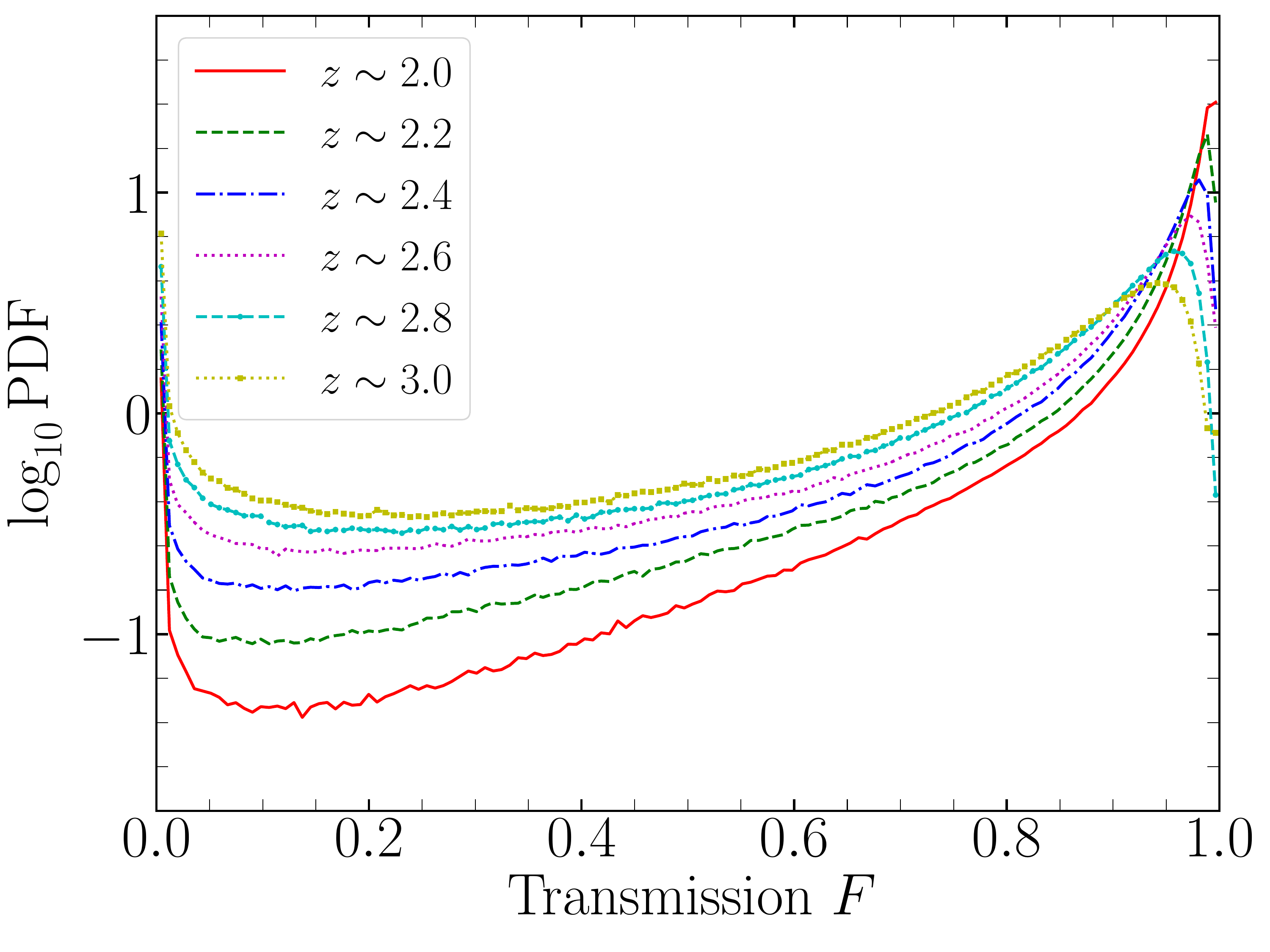}{0.45\textwidth}{(a)}
          \fig{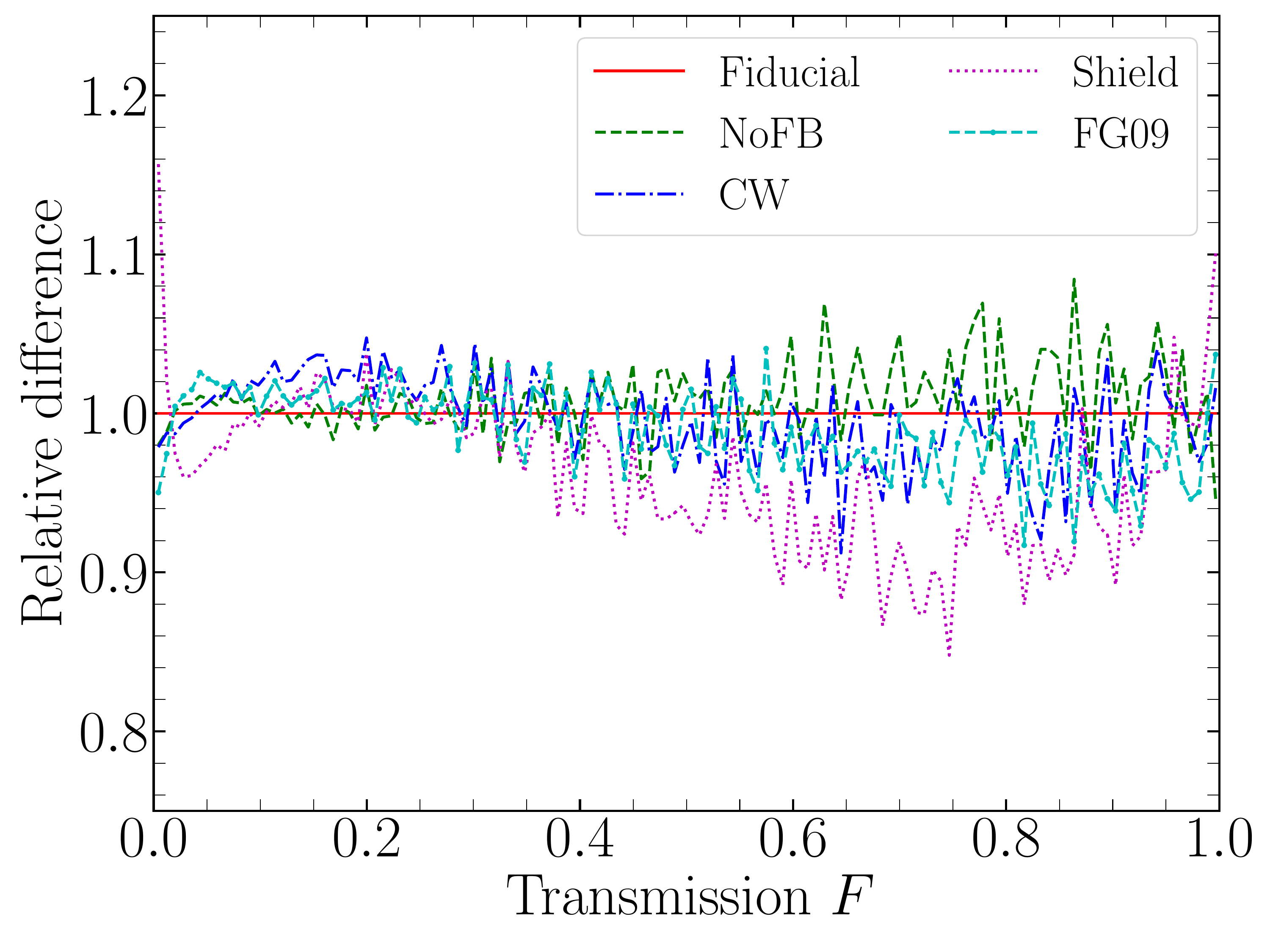}{0.45\textwidth}{(b)}}
\gridline{\fig{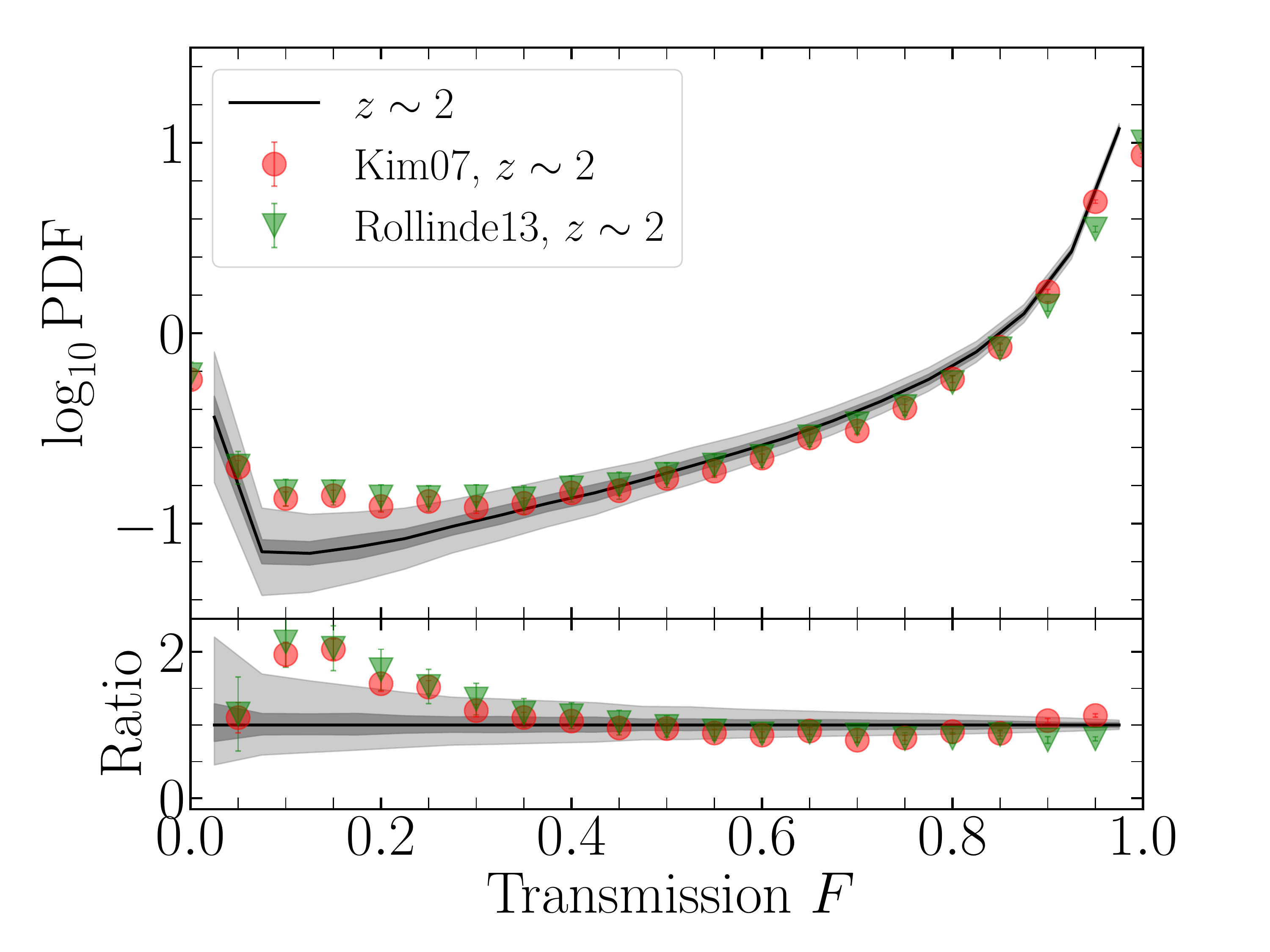}{0.45\textwidth}{(c)}
          \fig{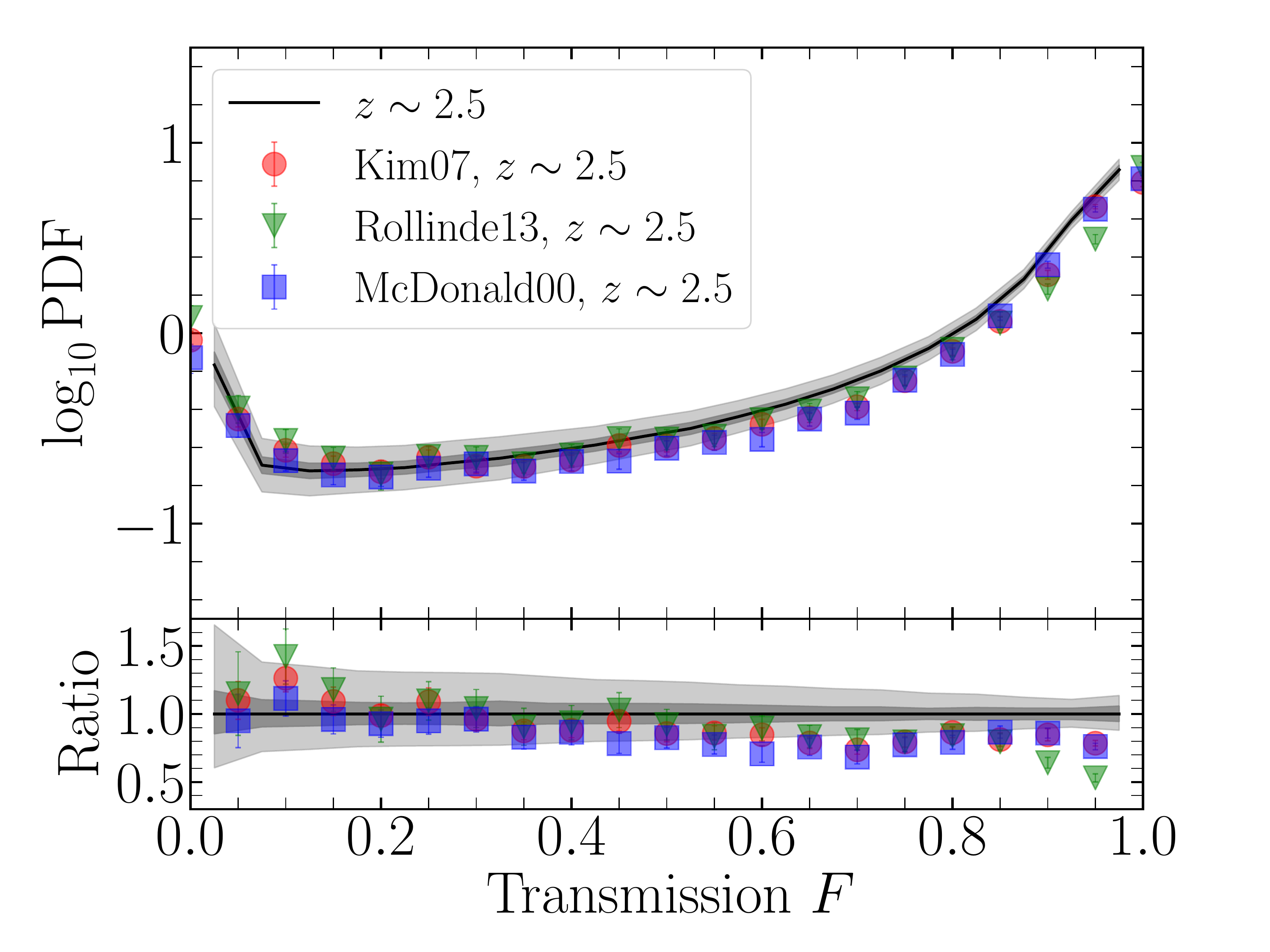}{0.45\textwidth}{(d)}}
\gridline{\fig{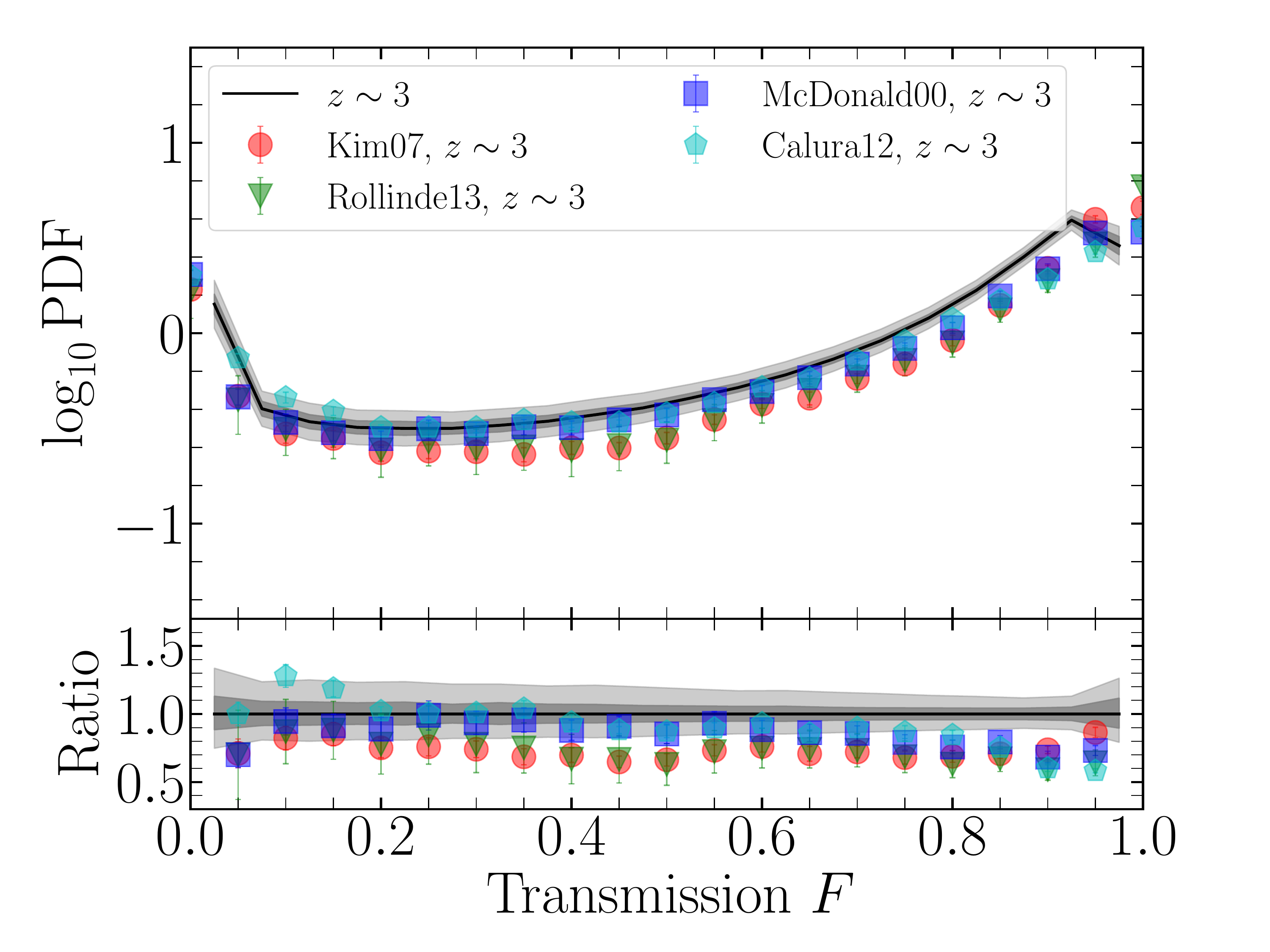}{0.45\textwidth}{(e)}}
 \caption{PDFs of the Ly$\alpha$ forest transmission flux of all LOSs from our fiducial run (Osaka20-Fiducial).  Panel (a) shows the evolution of PDF from $\bar{z}=2-3$ for the Osaka20-Fiducial run. Here, $\bar{z}\sim 2.0$ covers $1.9<z<2.1$, $\bar{z}\sim 2.2$ covers $2.1<z<2.3$, and so on.    
 Panel (b) shows the difference of PDFs for five different runs relative to the Fiducial run at $\bar{z}\sim 2.2$.
Panels (c)--(e) compare our results at $\bar{z}\sim 2-3$ against the observational data of \citet{McDonald00}, \citet{Kim07} and \citet{Rollinde13}. The black solid line is obtained by using the entire LOS data set as in the observations. The shaded regions are the 1$\sigma$ error estimated by the jackknife method (dark gray) and the 1$\sigma$ variance of all LOSs (light gray). The redshift coverage in panels (c)--(e) is the same as in \citet{Rollinde13}: $\bar{z}\sim 2$ ($1.88<z<2.37$), $\bar{z}\sim 2.5$ ($2.37<z<2.71$), $\bar{z}\sim 3$ ($2.71<z<3.21$). The bottom subpanels of (c)--(e) indicate the offset between the black solid line and the observational data points.  
  }
   \label{fig:fluxPDF}
\end{figure*}

%%%%%%%%%%%%%%%%%%%%%%%%%%%%%%%%%%%%%%%%%%

\section{Results}
\label{sec:result}

\subsection{Ly$\alpha$ forest flux PDF}
\label{sec:pdf}

As a starting point of our analysis, we show in Figure~\ref{fig:fluxPDF} the PDF of the $\Lya$ forest transmitted flux $F= \langle e^{-\tau} \rangle$. Here a higher value of $F$ means more transmission, i.e., less absorption with lower $\tau$. 
Here, all flux PDFs are computed as histograms using all the LOS data in the relevant redshift range, which is a standard procedure.

Panel (a) shows the redshift evolution of flux PDF at $\bar{z}=2-3$ with 128 bins in the range $0<F<1$. This is obviously a higher resolution than the observational data points, but we also would like to examine the higher-resolution PDF for the future.  
For example, the line for $\bar{z}\sim 2$ covers the redshift range of $z=1.9-2.1$ in the light-cone data.
As the effective optical depth $\tau_{\rm eff}(z) = -\ln \langle F(z) \rangle$ increases from $z=2$ to $3$ \citep{Becker13a}, the absorption lines with high $\tau$ increase, and the plateau at low $F$ gradually rises.  Here, we follow the standard practice of normalizing the simulated transmitted flux to $\tau_{\rm eff}(z)$ of \citet{Becker13a} in order to account for the uncertain UVB \citep[e.g.,][]{Viel13b}. 
The peak at high $F$ ($>0.9$) instead 
becomes less steep toward higher values of $F$ as the redshift increases. 

We also compare the results of different runs in Figure~\ref{fig:fluxPDF}(b) 
by showing the offset from the Fiducial run, again using the same 128 bins as in panel (a).  
We find that the overall shape of the PDF does not change very much, and the relative difference remains within $\sim 5$\% for most of the runs. However, the scatter becomes larger at higher $F$ values, and in particular for the ``Shield" run with 10\% deviation at $F>0.6$. 
The Shield run has higher {\HI} densities than the Fiducial run due to its self-shielding treatment, which explains the larger number of low-$F$ lines and hence fewer pixels at $F>0.6$. 

There have been somewhat mixed results over the years regarding the impact of feedback on flux PDF.  For example, \citet{Theuns02,Kollmeier06,Tepper13} suggested that the $\Lya$ forest statistics are not greatly affected by galactic wind because the hot supernova bubbles preferentially expand into the voids without affecting the filaments very much. 
\citet{Kollmeier03} examined the impact of galaxy photoionization, and concluded that it has only a small impact on the conditional mean flux decrement of $\Lya$ absorption.
\citet{Cen05} suggested that the volume filling factor of metal-enriched bubbles is a strong test for the strength of galactic wind feedback. 
Both \citet{Viel13b} and \citet{Chabanier20} argued that the AGN feedback can affect the flux PDF and 1D power spectrum significantly. 
However, \citet{Sorini20} showed that the impact of AGN feedback on the {\sc Hi} distribution around galaxies might not be so strong, statistically speaking. Our simulations show that SN feedback can have moderate impact ($\sim 5$\%) on the flux PDF. 

% Fig. 4
\begin{figure*}
\gridline{\fig{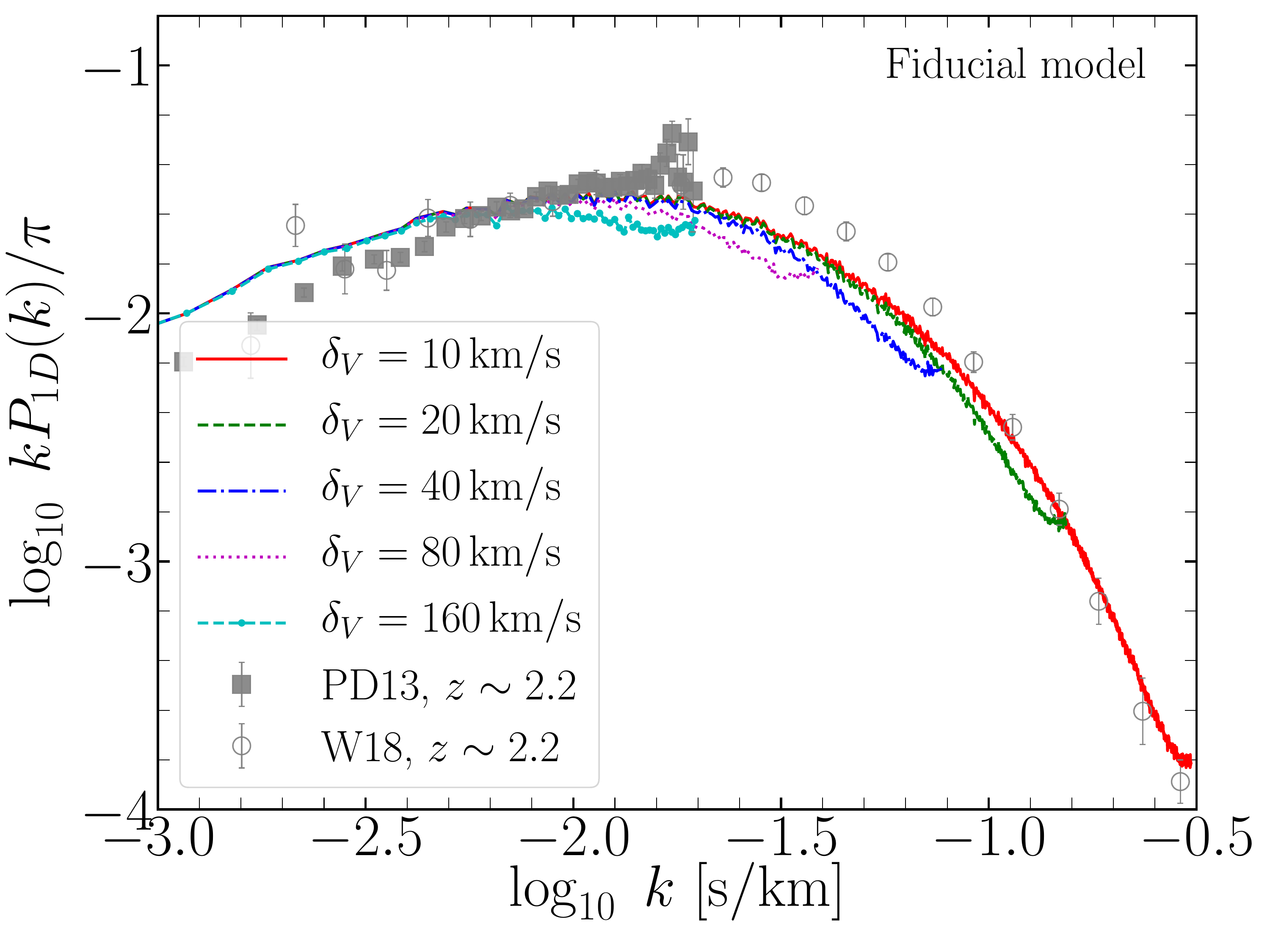}{0.47\textwidth}{(a)}
          \fig{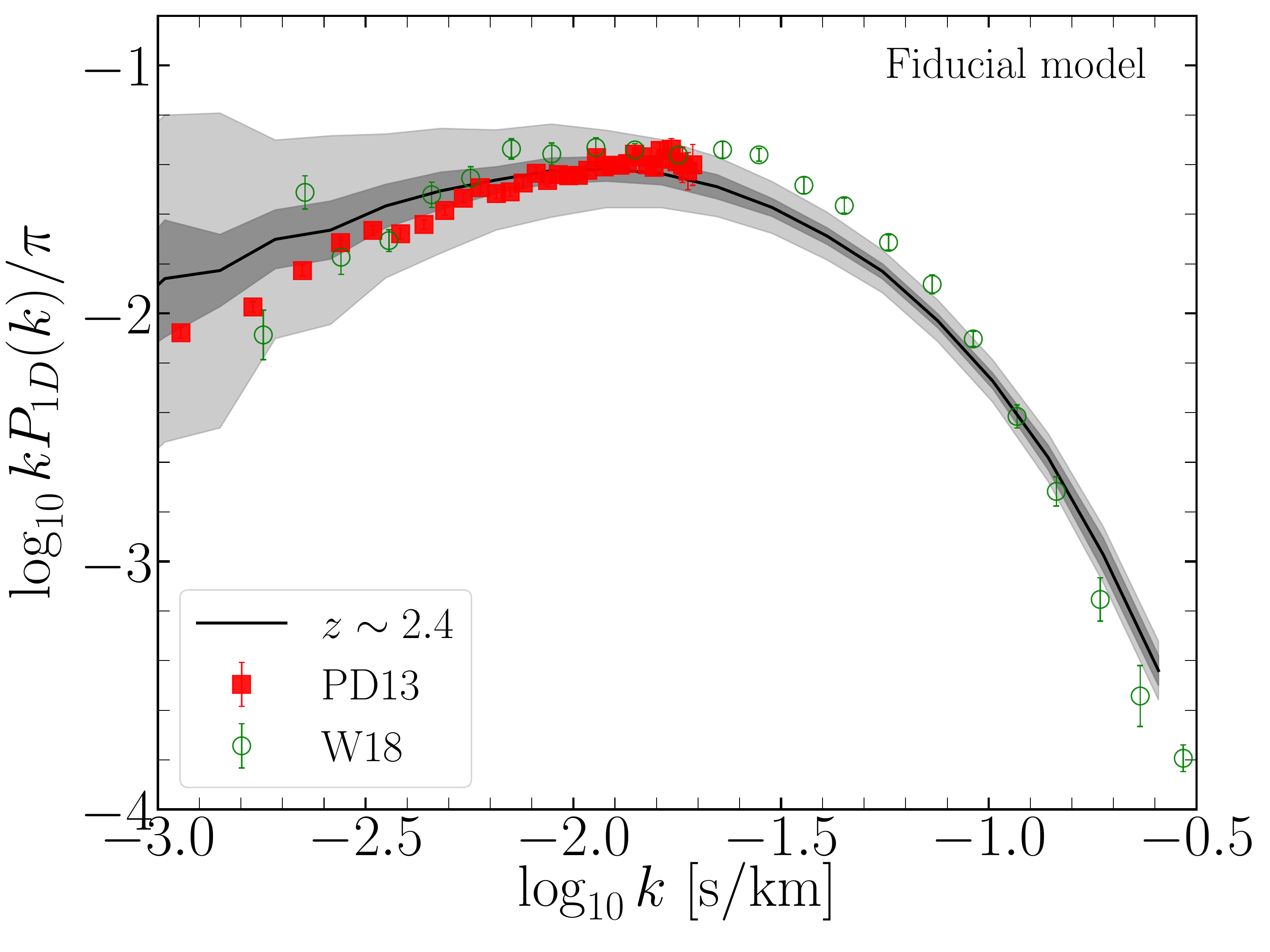}{0.47\textwidth}{(b)}}
\gridline{\fig{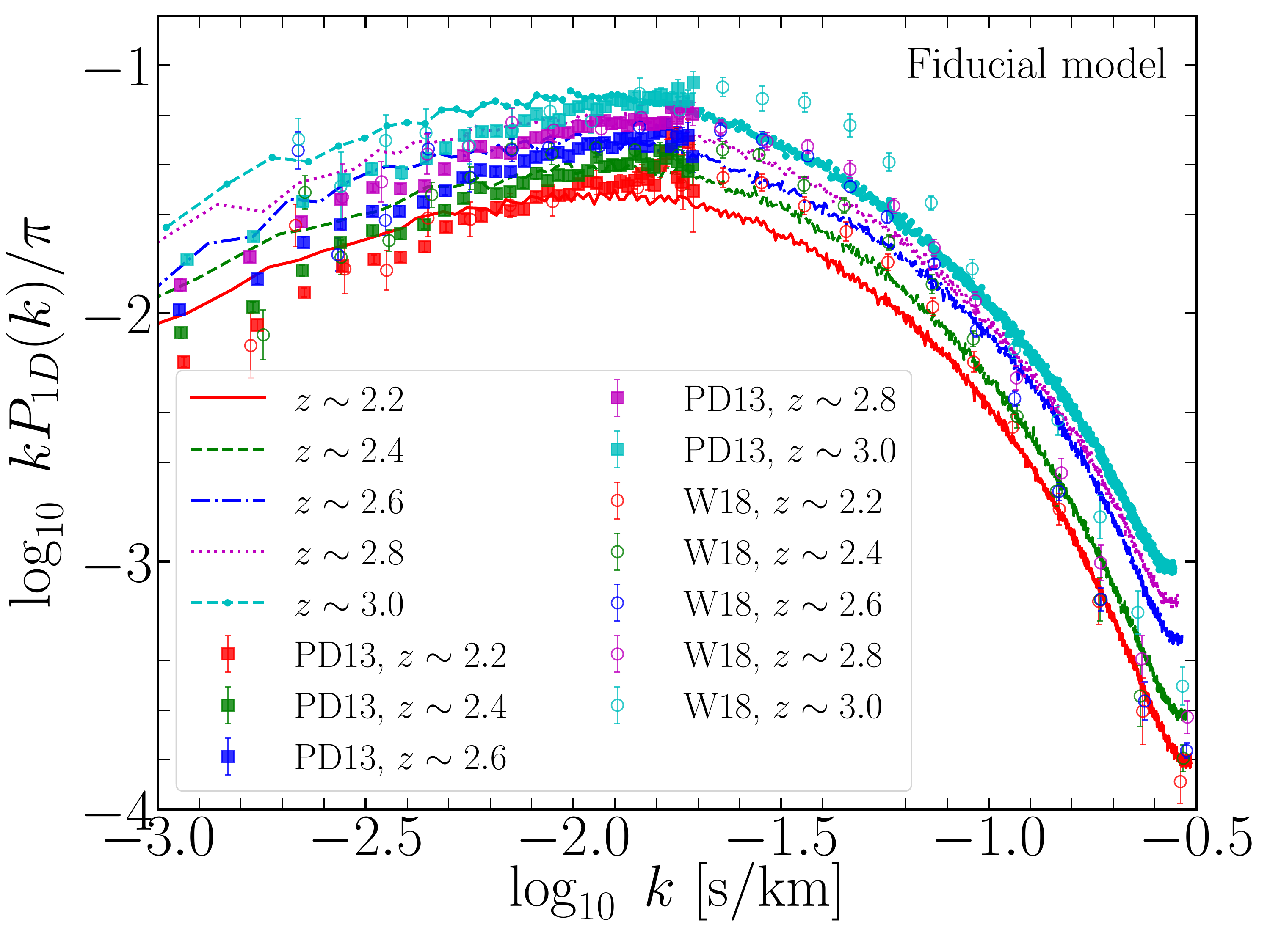}{0.47\textwidth}{(c)}
         \fig{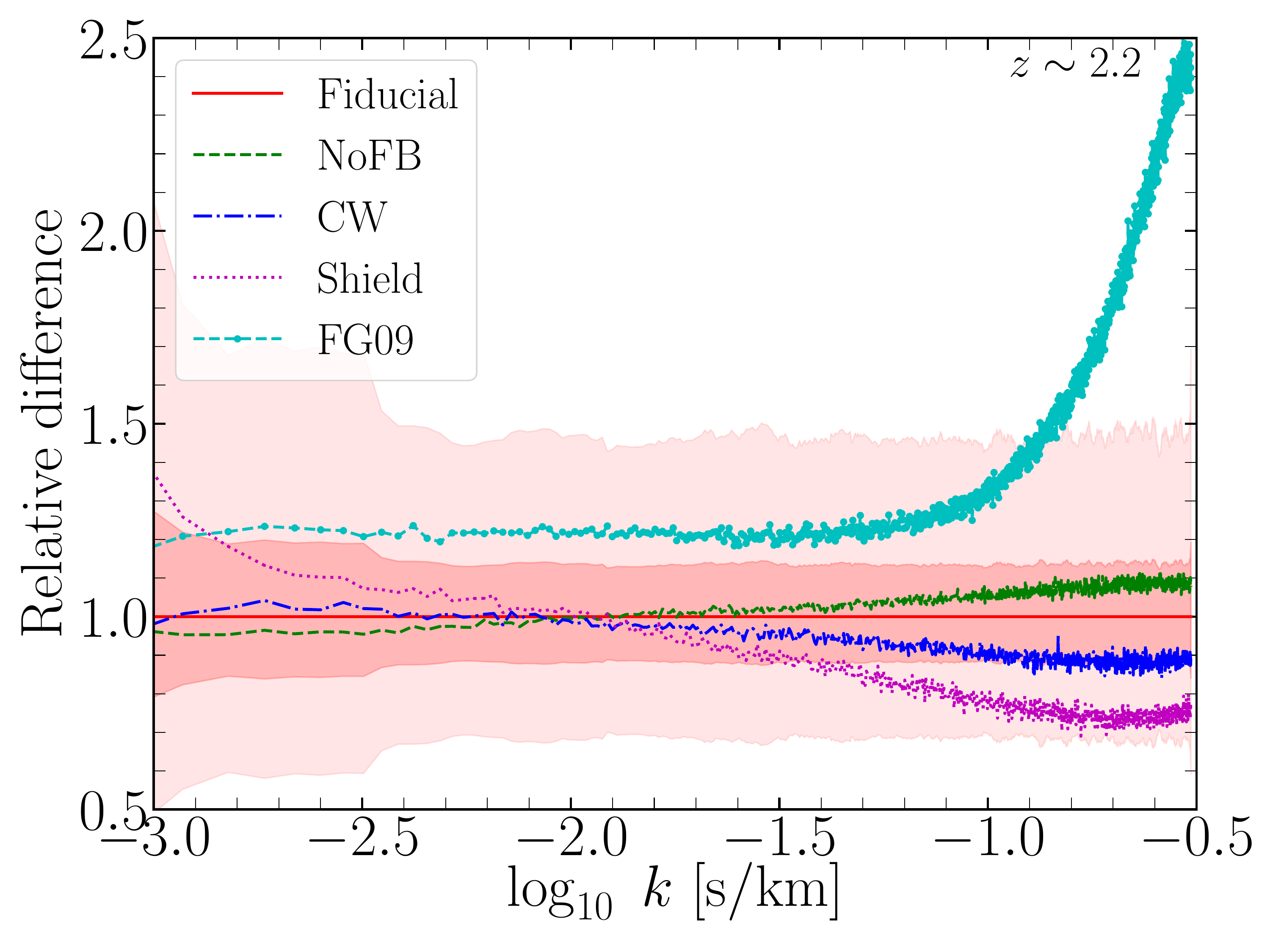}{0.47\textwidth}{(d)}}
\caption{1D power spectrum of the Ly$\alpha$ forest as a function of velocity wavenumber. 
    Panel (a) compares the $\Pk$ computed with different $\delta_v$ sizes for our Fiducial run at $\bar{z}\sim 2.2$.  As the value of $\delta_v$ becomes smaller, one can probe smaller scales more, and the upper end of $\Pk$ shifts to higher $k$ with decreasing $\delta_v$. In all panels, the observational data points are from \citet[][P13]{Palanque13} and \citet[][W18]{Walther18}.  
    Panel (b) shows the $\Pk$ with 1$\sigma$ jackknife error (dark gray) and 1$\sigma$ variance of all LOSs (light gray) in the Fiducial run using the same bins as W18 at $\bar{z}\sim 2.4$.
    Panel (c) shows the redshift evolution of $\Pk$ from $\bar{z}=2$ to 3 for the Fiducial model. 
    Panel (d) compares $\Pk$ of different simulation models given in Table~\ref{tab:sim} relative to the Fiducial run. The pink shades indicate 1$\sigma$ (darker one) and 3$\sigma$ (lighter one) jackknife errors.}  
    \label{fig:power}
\end{figure*}

In Figures~\ref{fig:fluxPDF}(c)--(e), we compare our result 
to the observational data points \citep{Kim07,Rollinde13} within the same redshift range as the observations.  \citet{Kim07} examined the flux PDF of 18 high-resolution quasar spectra observed with the Ultraviolet and Visual Echelle Spectrograph (UVES) on the Very Large Telescope (VLT) at $\bar{z}=2.07, 2.52,$ and 2.94 (from the LUQAS sample; \citealt{Kim04}). 
In their analysis, they separated the metal absorption lines and focused only on the {\HI} absorption. They found that the flux PDF is sensitive to the continuum fit in the range of $0.8<F<1.0$ and to metal absorption at $0.2<F<0.8$, where $F$ is the normalized flux. Kim et al.'s sample was larger than that of \citet{McDonald00}, and their measurement was systematically lower by up to 30\% at $0.2<F<0.8$, which can be seen in our Figure~\ref{fig:fluxPDF}(e). They ascribed this discrepancy due to a combination of improved removal of metal lines and cosmic variance. 
\citet{Rollinde13} analyzed four observational data sets: the LP sample (18 UVES VLT spectra from \citealt{Bergeron04}), the LUQAS sample, the \citet{Calura12} sample, and the \citet{McDonald00} sample. The sample covered the following redshift ranges: $\bar{z}=2$ ($1.88<z<2.37$), $\bar{z}=2.5$ ($2.37<z<2.71$), and $\bar{z}=3$ ($2.71<z<3.21$).
They compared their result to the mock sample from the GIMIC hydrodynamical simulation \citep{Crain09}, and found that the error estimated by the jackknife resampling method was smaller than the variance of the simulation result. 

In Figures~\ref{fig:fluxPDF}(c)--(e), the black solid line shows the result from all LOSs, the dark shade around it shows the 1$\sigma$ jackknife error obtained in a similar manner to the observations, and the wider light gray shade around it shows the 1$\sigma$ sample variance of all LOSs. We estimated the jackknife error by dividing the sample with 1024 LOSs into 64 subsamples, and estimated the variance by excluding one subsample each time (i.e., 64 trials). 
The jackknife errors are of similar size to those of the observed data points, but the sample variance of the simulation data is somewhat larger, which is consistent with the results of \citet{Rollinde13}. 
We see rough agreement between the simulation and observations, but also some deviations depending on the range of $F$ and redshift. For example, the simulation result is persistently higher than \citet{Kim07}'s data by more than 1$\sigma$ at $F>0.3$ at $\bar{z}\sim 3$.
The bottom subpanels of Figures~\ref{fig:fluxPDF}(c)--(e) show the ratio between the black solid line and the observed data points. At $\bar{z}\sim 2$, the deviation becomes somewhat large at $F<0.3$, but otherwise the simulation results and the data points agree with each other within 20\%--30\%. 
The discrepancy between observations and simulation result at the highest $F$ values is nearly universal among hydrodynamic simulations of the Ly$\alpha$ forest, and in fact likely due to systematic continuum-fitting errors in the observational data \citep{Lee12} rather than because of the simulations. 
\citet{Bolton17} found that the observational data points were within 2$\sigma$ of the flux PDF from the {\sc Sherwood} simulation, although in about half of the bins the observational data were below the simulation result by more than 1$\sigma$.  They suggested that their simulation might be lacking hot underdense gas because adopting an isothermal temperature--density relation improves the agreement at $0.1<F<0.8$.

\subsection{1D Ly$\alpha$ power spectrum}
\label{sec:Pk}

In addition to the flux PDF, the one-dimensional (1D) power spectrum of the $\Lya$ forest, $\Pk$, is another important observed statistic \citep{Croft98,McDonald06,Palanque13,Chabanier19}. 
In their pioneering work, \citet{Croft98} presented a simple method to recover the shape and amplitude of the power spectrum of matter fluctuations from observations of the $\Lya$ forest using a fast Fourier transform (FT) method. 
\citet{Palanque13} measured the $\Pk$ using 13,821 high-quality quasar spectra from SDSS-III/BOSS DR9 at $z=2.2-4.4$ on scales of $k=0.001-0.02$\,s\,km$^{-1}$. 
They improved the constraints on cosmological parameters from the SDSS by a factor of a few over the previous estimate by  \citet{McDonald06}, deriving $\sigma_8=0.83\pm 0.03$ and $n_s=0.97\pm 0.02$ assuming a flat-$\Lambda$ universe with no massive neutrinos, although without considering the astrophysical impacts of feedback in their hydrodynamic simulations. 
\citet{Viel13b} examined the $\Pk$ using the {\small OWLS} cosmological hydrodynamic simulations \citep{Schaye10}, and argued that the effect of galactic wind is comparable to the uncertainties of observed power spectra, and must be taken into account for a robust and accurate measurement. 

Here we follow Croft's FT method for our calculation of $\Pk$ for its simplicity and clarity. 
Figure~\ref{fig:power} presents $\Pk$ from our simulations compared with the observational data of \citet{Palanque13,Walther18}.
Some previous numerical works have added a Gaussian noise of $6-7\,\kms$ to the absorption profile data in order to mimic the observational error due to limited spectral resolution. 
Here, instead, we vary the velocity bin size when we compute the averaged absorption profile to evaluate the impact of spectral resolution. 
In Figure\,\ref{fig:power}(a), we show $\Pk$ for five different velocity bin sizes of $\delta_v=10, 20, 40, 80,$ and $160\,\kms$. 
As expected, with smaller spectral bin sizes, the $\Pk$ reaches higher $k$ and probes smaller scales.  
Note that $\delta_v = 160$\,km\,s$^{-1}$ corresponds to $\sim 2\,\hicmpc$ in the adopted cosmology. 
Our result with $\delta_v = 10$\,km\,s$^{-1}$ captures the upper end of $\Pk$ very well at high $k$, but some differences can be seen at the intermediate scale of $\log (k\,{\rm [s/km]}) \sim -1.5$ and at the largest scales of $\log (k\,{\rm [s/km]}) \lesssim -2.5$. 

In Figure\,\ref{fig:power}(b), we compare our simulation results to the data of \citet{Palanque13,Walther18} at $\bar{z}\sim 2.4$ (using the light-cone data at $z=2.3-2.5$).  
In this panel, we compute the 1$\sigma$ error (dark shade) by following the jackknife method \citep{Kim04,Lidz06,Rollinde13}: i.e., we divide the sample with 1024 LOSs into 64 subsamples, compute the $\Pk$ from each subsample, and then estimate the 1$\sigma$ error given by the subsamples within the same bins as W18 data points. 
Our jackknife errors are comparable to those on the W18 data points. 
Some of the deviations between simulation and the observed data points are within 1$\sigma$ error, but some are not.
In the same panel, we also show the 1$\sigma$ variance of all 1024 LOSs in light gray shade; it is larger than the 1$\sigma$ jackknife error.

Figure\,\ref{fig:power}(c) shows the redshift evolution  of $\Pk$ for the Fiducial run at $z\sim 2.2-3.0$. As the effective $\Lya$ optical depth increases toward higher redshift, the normalization of $\Pk$ also rises, and our simulation captures this evolution qualitatively well, with similar levels of differences from observational data as in Figures\,\ref{fig:power}(a) and (b). These differences from  the observational data are understandable because we have not fine-tuned our IGM thermal state by tweaking the parameters of  temperature--density relation ($T \propto \rho^\gamma$) as other simulation works have done assuming the FGPA 
\citep[e.g.,][]{Croft98,Weinberg98b,Sorini16}.

In Figure\,\ref{fig:power}(d), we compare simulations with different models listed in Table~\ref{tab:sim} against the Fiducial run at $\bar{z}\sim 2.2$. The same data points as in panel (a) are shown, as well as 1$\sigma$ and 3$\sigma$ jackknife errors in pink shades.  
We see that the FG09 run gives persistently higher power than the Fiducial run, which can be explained if the FG09 run has more {\HI} gas with less photoionization. In fact, this is consistent with the result by \citet{Rollinde13}, who derived a lower photoionization rate for the FG09 UVB model (see their Figure\,9). 
The NoFB run has a stronger power at higher $k$ range (but within 1$\sigma$), which can be explained by stronger clumping of {\HI} in high-density regions due to lack of SN feedback. It is not very clear why the Shield run has more (less) power at lower (higher) $k$ than the Fiducial run, and it deviates more than 1$\sigma$ at $\log (k\,[{\rm s/km}])<-2.8$ and $>-1.6$.

%%%%%%%%%%%%%%%%%%%%%%%%%%%%%%%%%%%%%%%%%%%%%%
\subsection{Flux Contrast vs. Impact Parameter}
\label{sec:contrast}

We define the Ly$\alpha$ ``flux contrast" $\eta_F$ as follows: 
\begin{equation}
    \eta_F = -\delta_F = 1 - \frac{F}{\langle F \rangle}, 
\label{eq:decrement}
\end{equation}
where $\delta_F$ is the absorption decrement integrated over a given velocity window within the
1D LOS, and  $F=\exp(-\tau)$ is the transmitted flux. 
The value of $\eta_F$ increases to unity with stronger absorption due to a greater amount of {\sc Hi}.
In the opposite limit of smaller $\tau$, $\eta_F$ approaches zero when $F$ approaches $\langle F \rangle$.

% Fig. 5
\begin{figure*}
\gridline{\fig{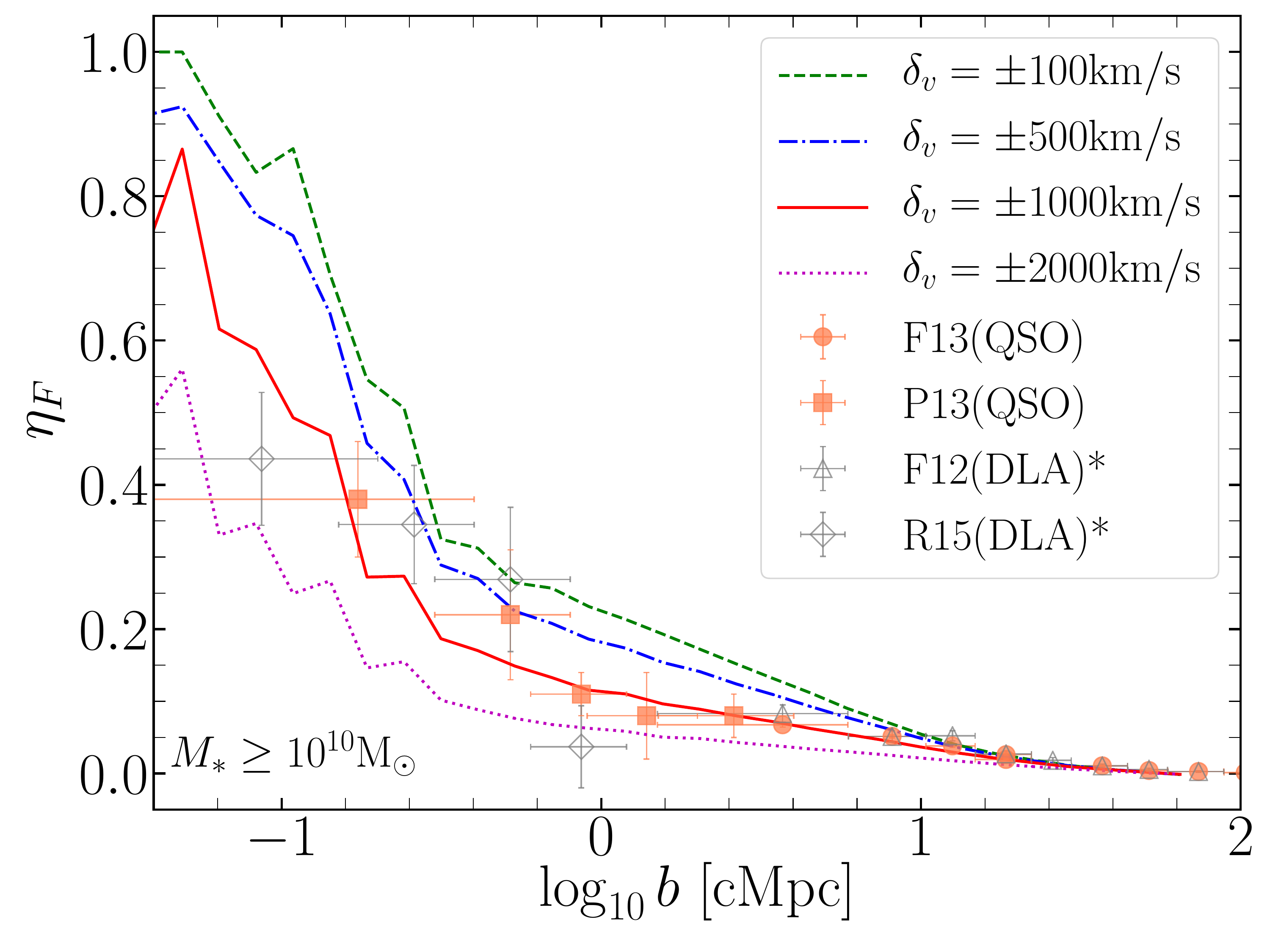}{0.43\textwidth}{(a)}
          \fig{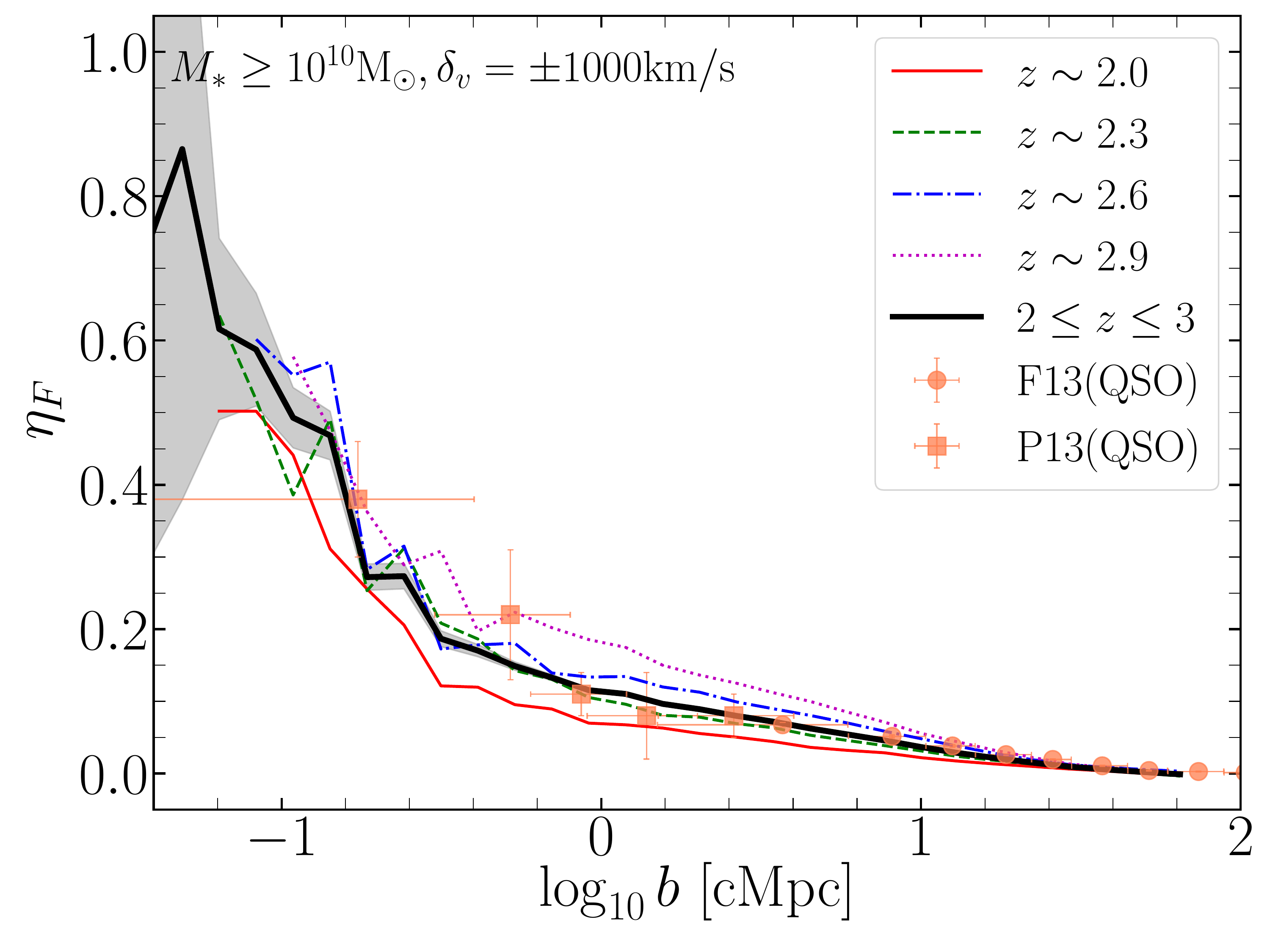}{0.43\textwidth}{(b)}}
\gridline{\fig{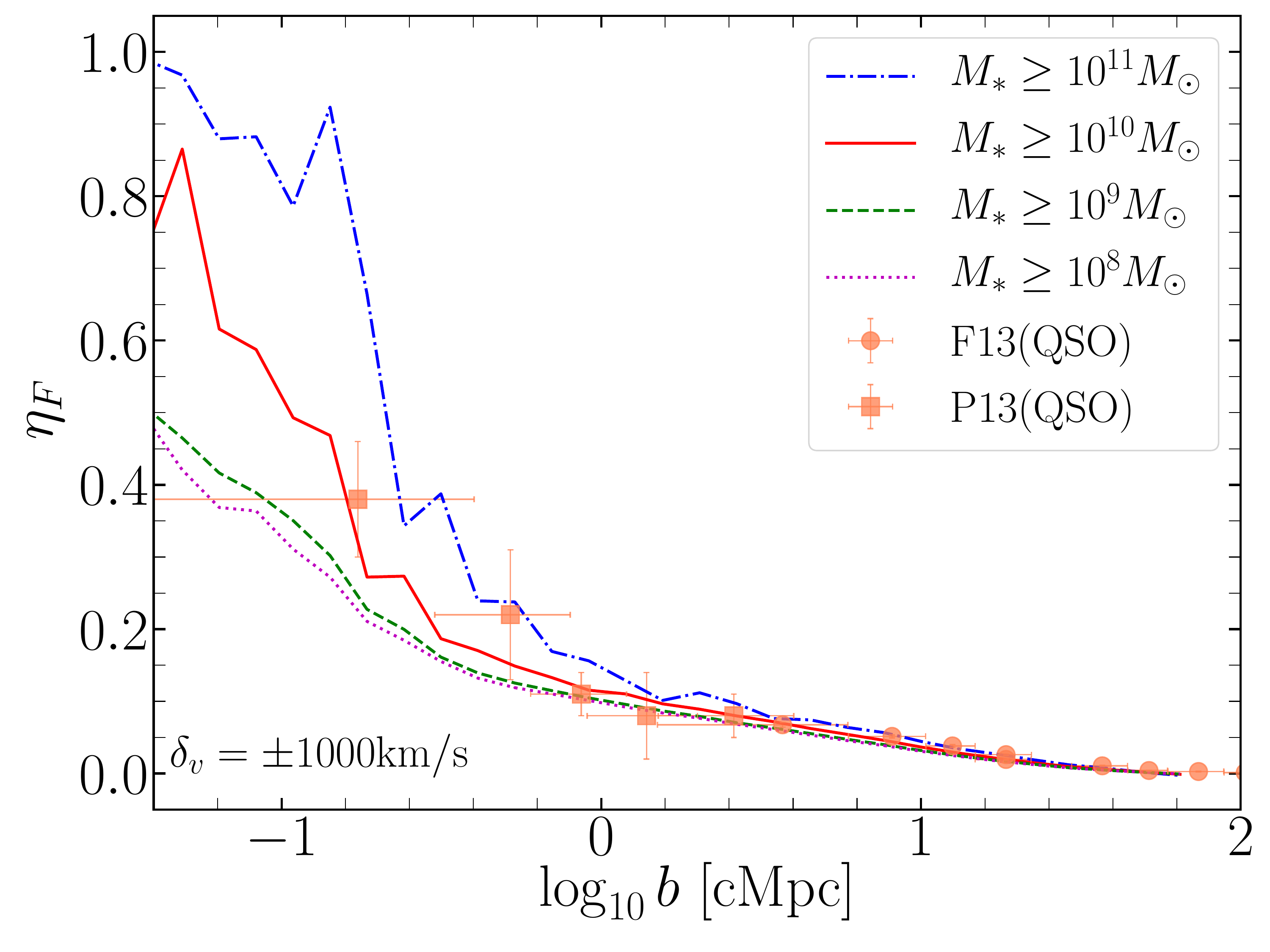}{0.43\textwidth}{(c)}
          \fig{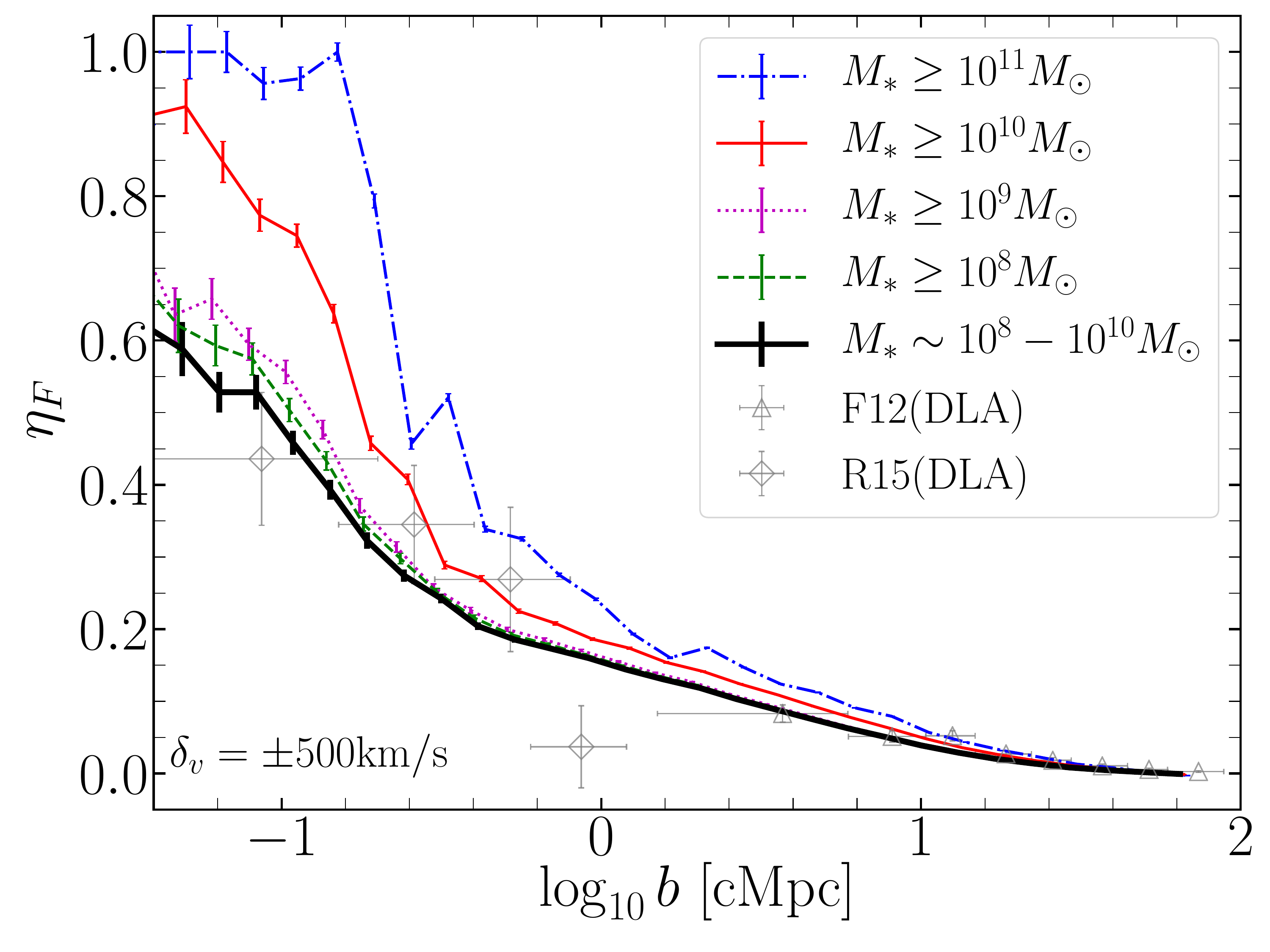}{0.43\textwidth}{(d)}}
\gridline{\fig{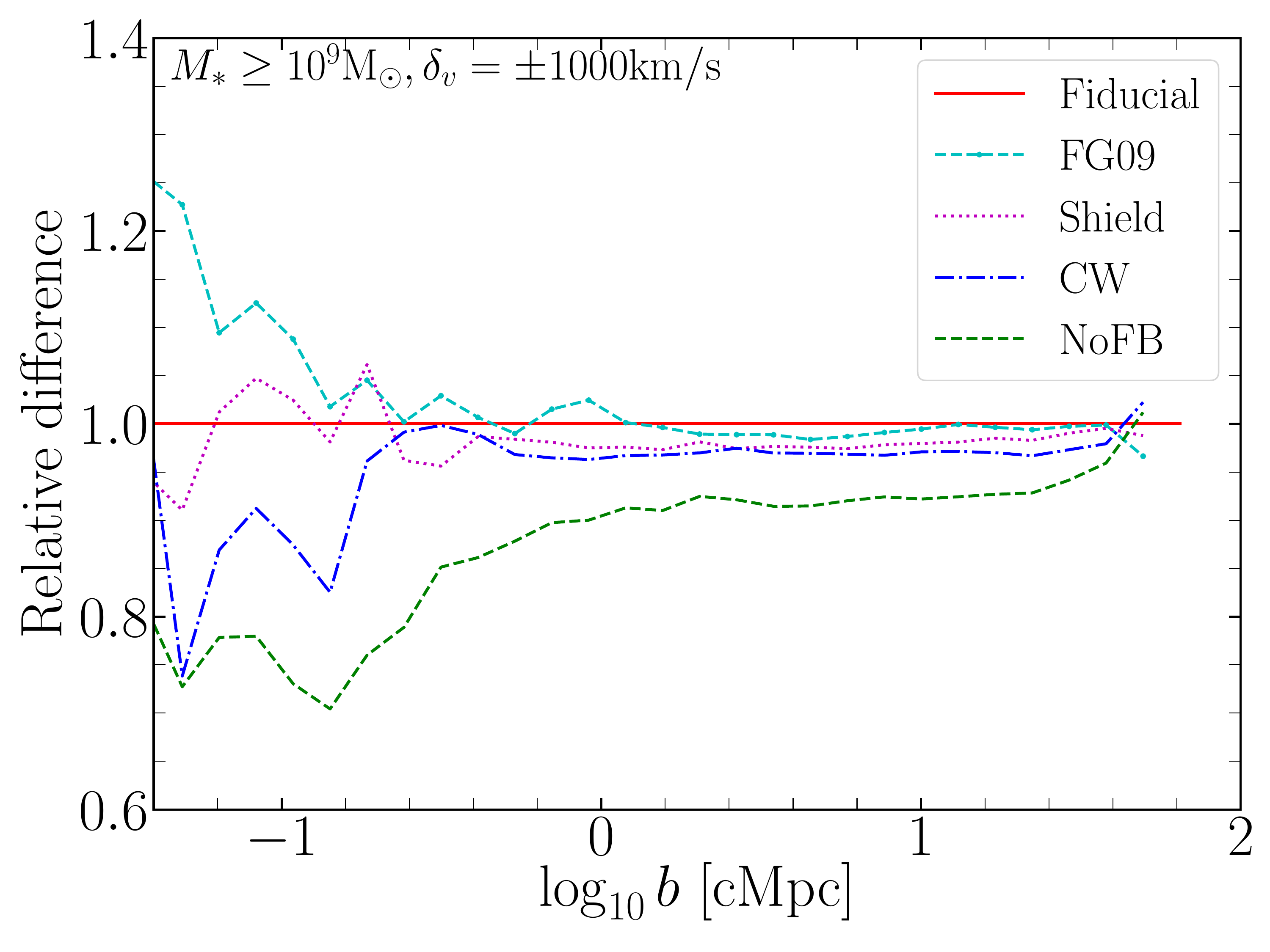}{0.43\textwidth}{(e)}
          \fig{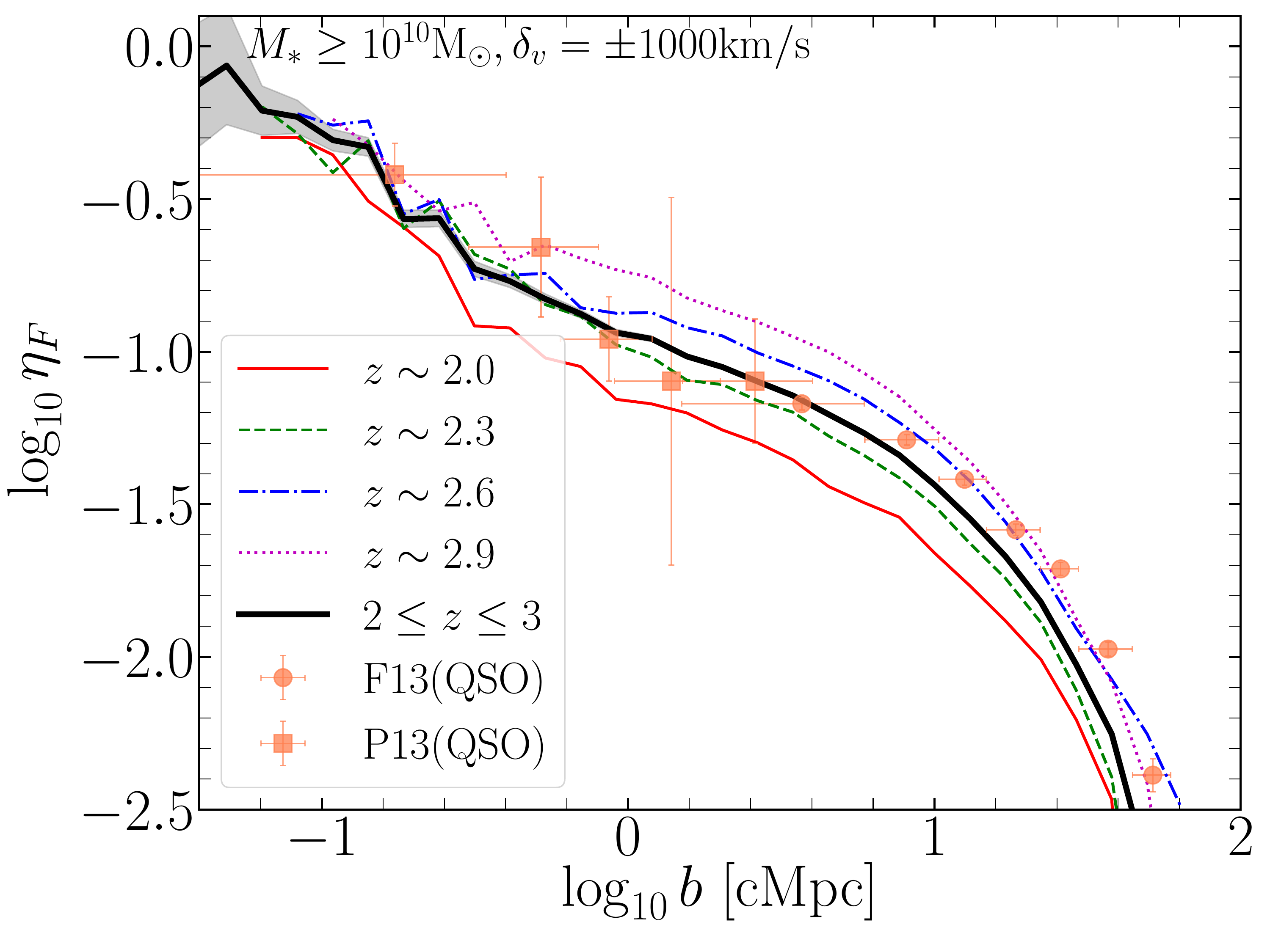}{0.43\textwidth}{(f)}}
    \caption{Flux contrast $\eta_F$ vs. impact parameter $b$ in units of comoving Mpc.  
    Panel (a) compares the results computed with different search ranges $\delta_v$.  Panel (b):  Redshift evolution of $\eta$ at $\bar{z}=2-3$.  The black solid line uses all the light-cone data, and the gray shade represents the error $\sigma/\sqrt{N}$, where $N$ is the data sample size in each bin. 
    Panels (c), (d): Dependence of $\eta_F$ on galaxy stellar mass. Here the data points with error bars are taken from \citet{Sorini18}, who converted the cross-correlation signal from the following authors to flux contrast:  \citet[][gray open triangle; F12]{Font-Ribera12}, \citet[][orange filled circle; F13]{Font-Ribera13}, \citet[][orange filled square; P13]{Pro13}, and \citet[][gray open diamond; R15]{Rubin15}. The error bars in panel (d) represent $
    \sigma/\sqrt{N}$, where $N$ is the data sample size in each bin.  
    Panel (e): Comparison of different simulations listed in Table~\ref{tab:sim} relative to the Fiducial run for the full light cone.   
    Panel (f): Logarithmic version of panel (b) to show the differences on large scales better. The gray shade on the black solid line is the same as in panel (b).  
    }
    \label{fig:impact}
\end{figure*}

In Figure~\ref{fig:impact}, we present $\eta_F$ computed for various galaxy samples as a function of impact parameter $b$\,[$\hicmpc$] from the galaxies (i.e., transverse distance).  
 We first sit on a galaxy, take cylindrical bins around it with a fixed logarithmic bin size (dividing the range $b =[10^{-2}, 50]\,\hicmpc$ into 32 equal-sized logarithmic bins), count all LOSs that come into the bin, and compute $\eta_F$. Each LOS can be associated with multiple galaxies. We use the full light-cone data and all galaxies within the volume, where the number of galaxies in each sample is $N_{\rm gal}= 3302, 48928, 195412, 277814$ for galaxy stellar mass-limited samples of $\Mstar > 10^{11}, 10^{10}, 10^9, 10^8\,\Msun$.  We choose to divide the galaxy samples based on their stellar mass rather than the dark matter halo mass, because in this way there is no ambiguity for substructures within massive halos. 
 In the future, when the full Subaru PFS data set becomes  available, there will be a large galaxy catalog with $\Mstar$ estimates, and we will be able to construct $\Mstar$-limited samples of galaxies for these studies. 
We find a general trend that $\eta_F$ increases with decreasing $b$, meaning that the amount of {\HI} increases toward the center of galaxies. 
This trend is consistent with the earlier simulation results \citep[e.g.,][]{Bruscoli03,Kollmeier03,Kollmeier06,Meiksin15,Turner17,Meiksin17,Sorini18}. 

In Figure~\ref{fig:impact}(a), we compare $\eta_F$ computed with different search ranges of $\delta_v=\pm 500, \pm 1000,$ \& $\pm 2000$\,km\,s$^{-1}$ along the LOS, and using the galaxy samples with stellar masses $\Mstar \ge 10^9\,\Msun$.  
The value of $\eta_F$ is expected to be higher (i.e., stronger  absorption) for smaller $\delta_v$ as it probes only the region closer to the galaxies. 
We compare our simulation results to the observational data points taken from \citet{Sorini18}, who converted the cross-correlation signal from the following authors to the flux contrast:  \citet[][grey triangle; F12]{Font-Ribera12}, \citet[][orange circle; F13]{Font-Ribera13}, \citet[][orange square; P13]{Pro13}, and \citet[][gray diamond; R15]{Rubin15}.
The data of F13 and P13 are for $\delta_v=\pm 1000$, and those of F12 \& R15 for $\delta_v=\pm 500$ (hence the asterisk in Figure~\ref{fig:impact}(a) to distinguish them). 

F12 measured the large-scale cross-correlation between the $\Lya$ forest and damped $\Lya$ systems (DLAs) using the ninth data release \citep[DR9;][]{Ahn12-DR9} of BOSS. 
They detected the cross-correlation signal on scales up to 40\,$\himpc$, and it is fitted well by the linear theory prediction of the CDM model with redshift distortions. The amplitude of the DLA--$\Lya$ cross-correlation depends on the bias factor of the DLA system as well as that of the $\Lya$ forest. 

Using a similar technique to F12, F13 also measured the cross-correlation between quasars and the $\Lya$ forest in the redshift space using 60,000 quasar spectra from the BOSS DR9. They found that the cross-correlation can be fit well by the linear theory prediction at $b>15\hicmpc$, and that the quasar bias is $3.64^{+0.13}_{-0.15}$ at $\bar{z}=2.38$. They also argued that the failure of simple linear model at $b<15\himpc$ could be due to enhanced ionization by the quasar radiation, i.e., the quasar proximity effect \citep[e.g.,][]{Bajtlik88,Scott00,Calverley11}. 

Projected pairs of quasar (or QSO) sightlines allow us to study a foreground quasar's environment in absorption lines.  
P13 used the LOSs of 650 quasar pairs to study the {\HI} absorption transverse to luminous $z\sim 2$ quasars at proper separations of 30\,kpc$< b <$ 1\,Mpc. Their analysis of composite spectra revealed excess $\Lya$ absorption characterized by $W= 2.3{\rm \AA} (b/100\,{\rm kpc})^{-0.46}$. The excess of optically thick {\HI} absorbers ($\NHI>10^{17.3}$\,cm$^{-2}$) at $b<200$\,kpc was described by a quasar--absorber cross-correlation function $\xi_{\rm QA}=(r/r_0)^\gamma$ with $r_0=12.5^{+2.7}_{-1.4}\hicmpc$ and $\gamma=1.68^{0.14}_{-0.30}$, which is consistent with quasars being hosted by massive dark matter halos with $\Mh \sim 10^{12.5}\Msun$ at $z\sim 2.5$ according to P13.

Similarly to P13, R15 also used close pairs of quasars to probe the CGM transverse to 40 DLAs at $1.6<z_{\rm DLA}<3.6$.  From the analysis of average $\Lya$ absorption profiles, they showed that the covering fraction of optically thick {\HI} ($\NHI > 10^{17.2}$\,cm$^{-2}$) is $\gtrsim 30$\% within $b<200$\,kpc, and that of {\sc Si ii} is $20^{+12}_{-8}$\% within $b<100$\,kpc.

In Figure~\ref{fig:impact}(a), we see rough agreement with observational data for $\delta_v = \pm 1000$\,km\,s$^{-1}$ within the error bars, and the agreement on large scales is particularly good as it converges to zero toward the edge of the simulation box. This convergence is natural because the mean transmission is rescaled to match  $\langle F \rangle$ as we explained in Section~\ref{sec:pdf}.

Figure~\ref{fig:impact}(b) shows the redshift evolution of $\eta_F$ as a function of impact parameter $b$ for the Fiducial model. 
As the redshift becomes higher, the number of more massive galaxies decreases, and the signal on small scales becomes somewhat noisy at $\log (b\,[{\rm cMpc}])\lesssim -1.0$. Therefore we have truncated the lines at the smallest scale bins where we cannot make reliable measurements due to small statistics. 
In this panel we are using the galaxy sample with $\Mstar > 10^{10}\,\Msun$ to compare with the data of F13 and P13 which are based on quasar sightlines. Generally quasars are considered to reside in the most massive halos of $\Mh > 10^{12}\,\Msun$, therefore taking the galaxy sample with $\Mstar > 10^{10}\,\Msun$ would roughly correspond to such halos and be more appropriate. 
The logarithmic version of Figure~\ref{fig:impact}(b) is shown in Figure~\ref{fig:impact}(f), which shows the behavior of different models more clearly on large scales. A nice agreement within the error bars can be seen at $\log (b\,[{\rm cMpc}])\lesssim 1.0$.  The data points are somewhat on the higher side at $\log (b\,[{\rm cMpc}])\gtrsim 1.2$, but since the ordinate is on a logarithmic scale, the actual deviation is small as seen in panels (a) and (b).      

Figures~\ref{fig:impact}(c) and (d) show $\eta_F$ for different galaxy samples with stellar mass ranges of $\Mstar \ge 10^8, 10^9, 10^{10}$, and $10^{11}\,\Msun$.  We find higher $\eta_F$ values for more massive galaxies, as expected from the higher bias of massive galaxies in more massive halos. 
Figure~\ref{fig:impact}(c) compares the simulation results to the data of F13 \& P13 with $\delta_v = \pm 1000$\,km\,s$^{-1}$, and Figure~\ref{fig:impact}(d) to the data of F12 \& R15 with $\delta_v = \pm 500$\,km\,s$^{-1}$.
The lines for $\Mstar>10^{10}\,\Msun$ and $>10^{11}\,\Msun$ in Figure~\ref{fig:impact}(c) show a rough agreement with F13 \& P13 data, and this is reasonable for the QSO sightlines which would correspond to more massive halos with $\Mh \gtrsim 10^{12}\,\Msun$.
In Figure~\ref{fig:impact}(d) we see better agreement with lower $\Mstar$ samples with F12 \& R15 data points based on DLAs, which is also reasonable because median DLA halos are expected to have somewhat lower mass halos of $\Mh \lesssim 10^{12}\,\Msun$ \citep[e.g.,][]{Nag07a,Font-Ribera12}. 
Here the error bars on the simulation result are computed as $\sigma/\sqrt{N}$ in each bin, where $N$ is the number of LOSs that came into the radial bin.  
In this panel we also show the line for the sample with $10^8\,\Msun <\Mstar<10^{10}\,\Msun$ (thick black line), which is lower than those for $\Mstar>10^{10}\,\Msun$ or $>10^{11}\,\Msun$. This suggests that the $\eta_F$ signal can be boosted strongly by the massive galaxies. 
In other words, the data points of F12 and R15 contain important contributions from lower-mass galaxies with $\Mstar \le 10^{10}\,\Msun$ rather than just from a limited number of massive galaxies. 

The dependence of $\eta_F$ signal on the galaxy sample means that $\eta_F$ may not be converged yet with the limited box sizes of our simulations (see also Appendix~\ref{app:res}).  We caution the reader that the results of $\eta_F$ could depend strongly on the box size and associated galaxy mass range, and our results may not have fully converged yet. 

Figure~\ref{fig:impact}(e) shows the results of $\eta_F$ for different models and the simulation runs given in Table~\ref{tab:sim}.  This panel is computed with the galaxy sample of $\Mstar \ge 10^9\,\Msun$ and $\delta_v \ge \pm 1000$\,km\,s$^{-1}$.
We find that the variation of $\eta_F$ due to the model differences is less than 25\%, which is somewhat smaller than those due to $\delta_v$ or $\Mstar$ dependence.  All runs other than the NoFB run are consistent with the Fiducial run within 5\%  at $\log (b\,[{\rm cMpc}]) \gtrsim -0.4$.  
To further interpret the model dependences, one should consider the impact of star formation.  For example, one might naively expect that there will be more {\sc Hi} gas around galaxies for the NoFB run. However, it is known that galaxies become more dense and compact without galactic wind feedback \citep[e.g.,][]{Yajima17}, and the dense gas in  galaxies is vigorously converted into stars, which leads to the overprediction of galaxy stellar masses. As a result of this compact {\HI} distribution and overconsumption of {\HI} gas into stars, the NoFB run has the lowest $\eta_F$ near the galaxies among all models. In the CW run, the gas is ejected out of the galactic potential more efficiently than other models with some overheating of the IGM \citep{Choi11}, hence there is less {\HI} gas around galaxies than in the Fiducial run.  
These results indicate that it is crucial to treat star formation and feedback in simulations to obtain accurate results of $\eta_F$ at $b\lesssim 1$\,cMpc, and this statistic can be an important test of feedback models. 
The Shield run has a higher $\eta_F$ than the Fiducial run, up to 5\% in a few bins on small scales, and these could be the dense cold ISM in the substructure galaxies near the central galaxy, which could be more prominent due to the self-shielding effect. The FG09 run also has a higher $\eta_F$ than the Fiducial run by more than 10\% at the smallest scales of $b\lesssim 10$\,ckpc, which can be ascribed to lower photoionization rates than in the HM12 model (see also Sec.~\ref{sec:Pk} and Figure~\ref{fig:power}(d)).

%%%%%%%%%%%%%%%%%%%%%%%%%%%%%%%%%%%%%%%%%%

\subsection{Cross-correlation between galaxies and {\sc Hi}}
\label{sec:ccf}

%%Fig. 6
\begin{figure}
    \epsscale{1.2}
	\plotone{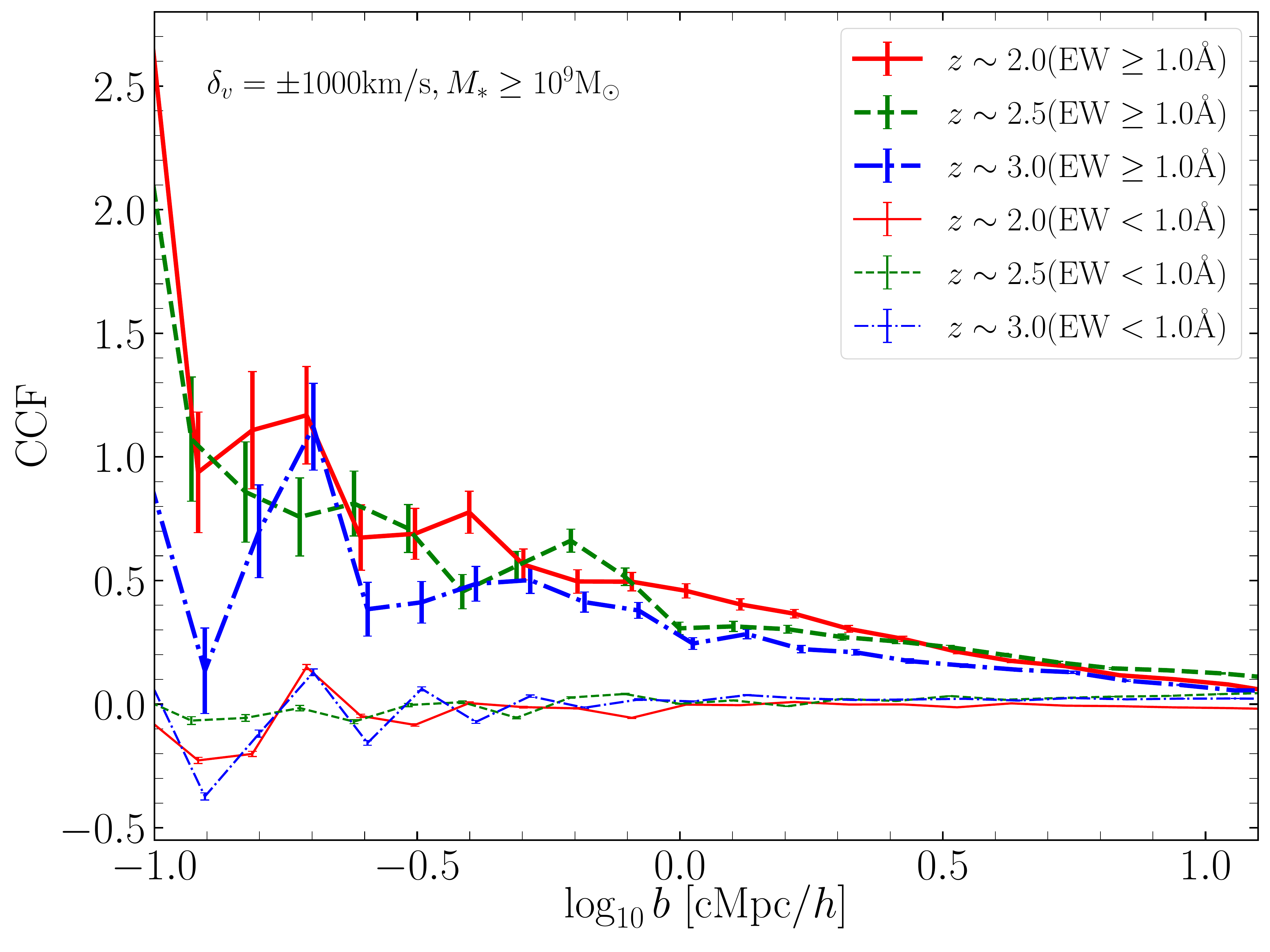}
    \caption{
    CCF between galaxies ($\Mstar \ge 10^{9}\,\Msun$) and {\sc Hi} absorbers with high EW (EW$>$1\,\AA; thick lines) and low EW (EW$<$1; thin lines) as a function of transverse distance between galaxies and LOS.  High-EW systems have stronger correlations with galaxies, while low-EW systems are more broadly distributed and hence have almost no correlation or negative correlation. See the main text for more details. 
    }
    \label{fig:ccf2DHI_gal}
\end{figure}

We use the same LOS and galaxy samples to compute the cross-correlation function (CCF) in the following. We first sit on a galaxy, take cylindrical bins around it with equal  logarithmic bins (32 logarithmic bins for the range of $b=[0.01, 20]\,\hicmpc$), count all LOSs that come into the bin, and compute the CCF between the galaxy and the pixels along the LOSs. Each LOS can be associated with multiple galaxies.  

Figure~\ref{fig:ccf2DHI_gal} shows the CCF between the simulated galaxies with $\Mstar \ge 10^{9}\,\Msun$ and discrete {\HI} absorbers with high equivalent width (EW) (EW$>$1\,\AA; thick lines) and low EW (EW$<$1; thin lines) as a function of transverse distance between galaxies and LOS. 
As for the CCF, we use a simple estimator of $(DD/DR) - 1$, where $DD$ and $DR$ are the numbers of pairs found in data--data and data--random data sets.  
Here the EW is simply measured as a boxcar EW. 
A more accurate treatment requires a photoionization calculation using a code such as Cloudy, and measurement of the {\sc Hi} column density via Voigt profile fitting.  
We plan to present such analyses together with metal absorption lines in the future. 
Here, for a simple examination of the dependence of the CCF on the {\sc Hi} column density, we divide the sample into two samples of high- and low-EW systems, and compute the CCFs separately at each redshift. 
In other words, our analysis presented here is performed with unsmoothed, finely sampled {\sc Hi} absorption lines compared to the actual observations.  

The high-EW systems have stronger correlations with galaxies as expected, while the low-EW systems are more broadly distributed in space and hence have almost no correlation (or even slightly negative correlation on small scales) relative to the galaxies. 
Note that the ordinate is on a linear scale in this plot to show the negative correlation at the same time. 
A weak redshift evolution is visible in  Figure~\ref{fig:ccf2DHI_gal}, with stronger correlations at $z=2$ than at $z=3$ due to the evolution in $\tau_{\rm eff}$ \citep{Becker13a}.  
The Poisson error bars are computed as $(1+{\rm CCF})/\sqrt{D_{\rm gal} D_{\rm abs}}$, where $D_{\rm gal}$ and  $D_{\rm abs}$ are the numbers of galaxy pairs and absorber pairs in each bin, respectively.

%%Fig. 7
\begin{figure}
    \epsscale{1.2}
	\plotone{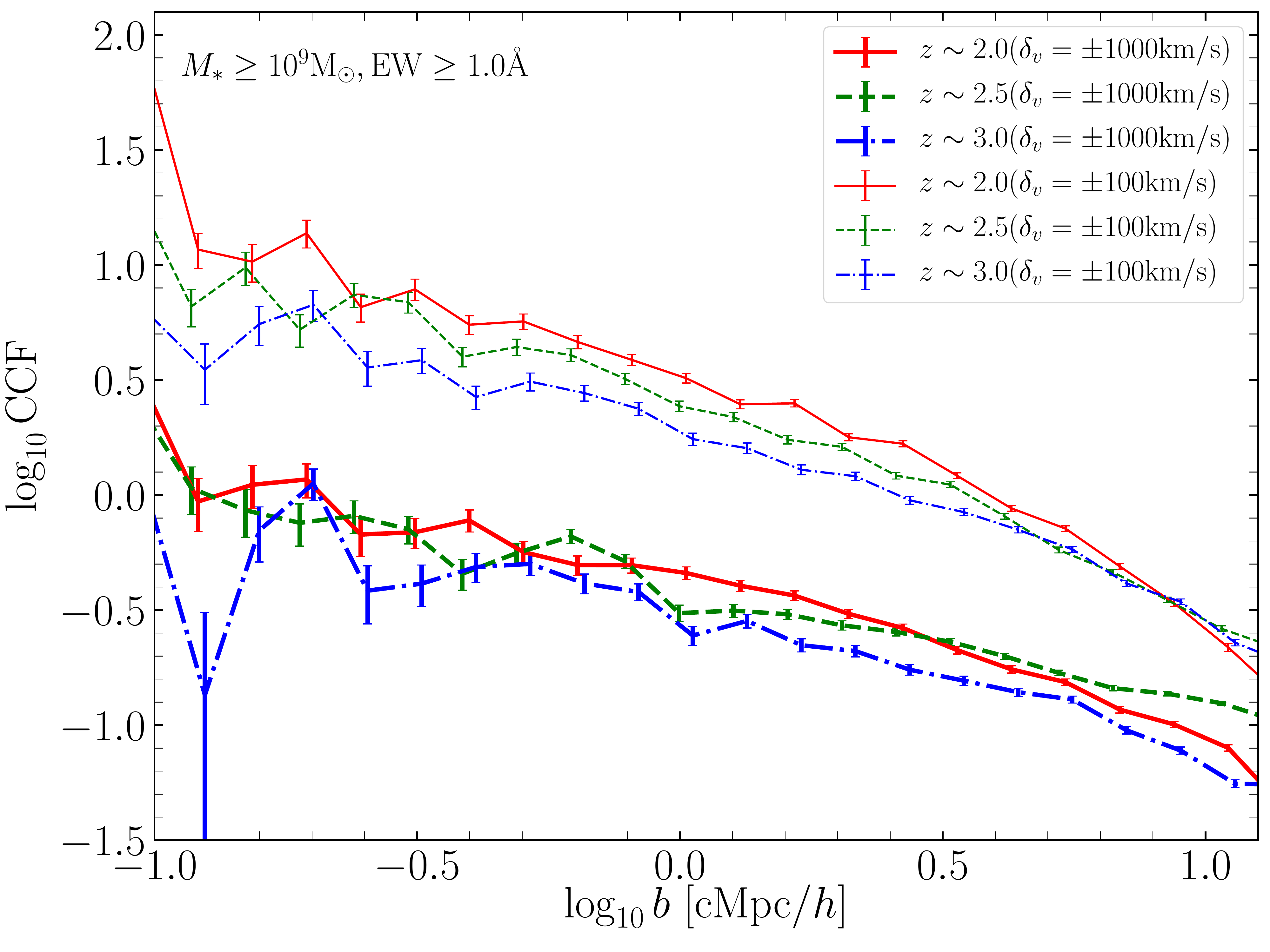}
    \caption{
    Dependence of the CCF on the search range along the LOS. 
    The thick lines are for search path length of $\delta_v = \pm 1000$\,km\,s$^{-1}$, and thin lines are for $\delta_v = \pm 100$\,km\,s$^{-1}$.  The latter case is only looking at correlations between close pairs of absorbers and galaxies, and hence produces a stronger CCF signal. 
    For all cases, galaxies with $\Mstar > 10^{9}\,\Msun$ and {\sc Hi} absorbers with high EW (EW$>$1\,\AA) were used.
    }
\label{fig:ccfzevol}
\end{figure}

When computing the pairs for the CCF, it is not so obvious how deeply one should search for those pairs along the LOS. 
In order to examine the impact of search range $\delta_v$, we repeat the CCF calculation with two different values of $\delta_v= \pm 1000$ \& $\pm 100$\,km\,s$^{-1}$, and show the results in  Figure~\ref{fig:ccfzevol}.  We first sit on a galaxy, and search for the \HI\ absorbers within $\pm \delta_v$ from the systemic velocity of the galaxy along the LOS.
The thick and thin lines are with $\delta_v = \pm 1000$\,km\,s$^{-1}$ and $\pm 100$\,km\,s$^{-1}$, respectively.  
The latter case is only looking at the correlations between close pairs of absorbers and galaxies, and hence produces stronger CCF signals. 
For all cases, galaxies with $\Mstar > 10^{9}\,\Msun$ and the high-EW absorber sample were used.

%%Fig. 8
\begin{figure}
    \epsscale{1.2}
	\plotone{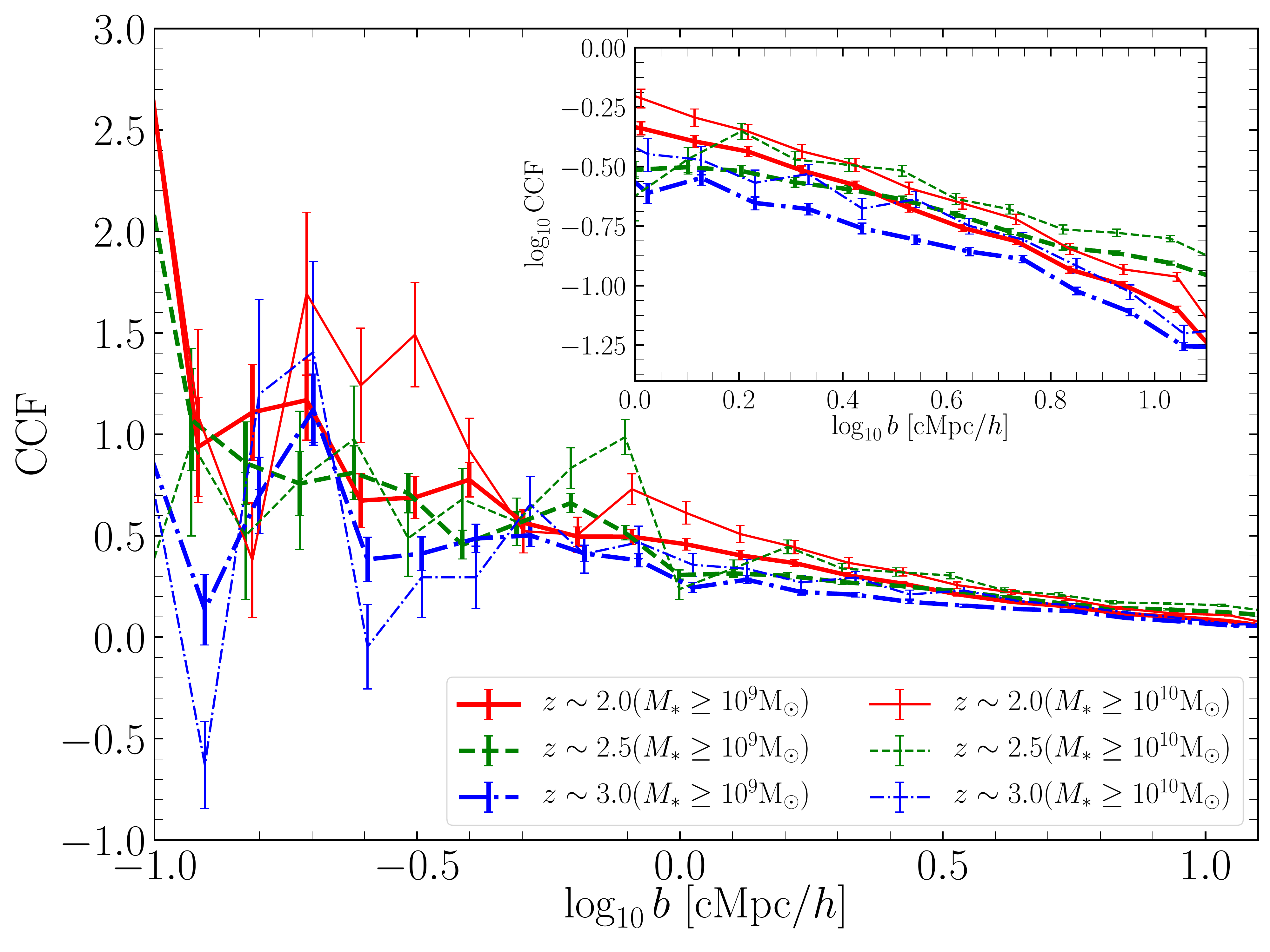}
    \caption{
   Dependence of the CCF on the stellar mass of galaxies of concern.  
   The thick lines are for galaxies with $\Mstar > 10^{9}\,\Msun$, 
   and thin lines are for those with $\Mstar > 10^{10}\,\Msun$.
   For all cases, {\sc Hi} absorbers with high EW (EW$>$1\,\AA) and the search range of $\delta_v = \pm 1000$\,km\,s$^{-1}$ were used.  
    }
\label{fig:ccfmass}
\end{figure}

In Figure~\ref{fig:ccfmass}, we show the dependence of the CCF signal on galaxy stellar mass. 
The number of galaxies with $\Mstar > 10^{10}\,\Msun$ becomes smaller at $z=3$ in our simulation, and the number of CCF pairs also becomes less, resulting in the somewhat noisy CCF signal on small scales. 
(For example, there are 3515 galaxies in the range $2.9<z<3.1$, which corresponds to two simulation snapshots of a $100\hicmpc$ box.) 
At least, we see that the overall CCF signal is stronger for the $\Mstar > 10^{10}\,\Msun$ sample (thin lines) than for the $\Mstar > 10^{9}\,\Msun$ sample (thick lines). 
This is consistent with the stellar-mass trend that we saw in Figure~\ref{fig:impact}(c) and (d). 

\begin{figure}
    \epsscale{1.2}
        \plotone{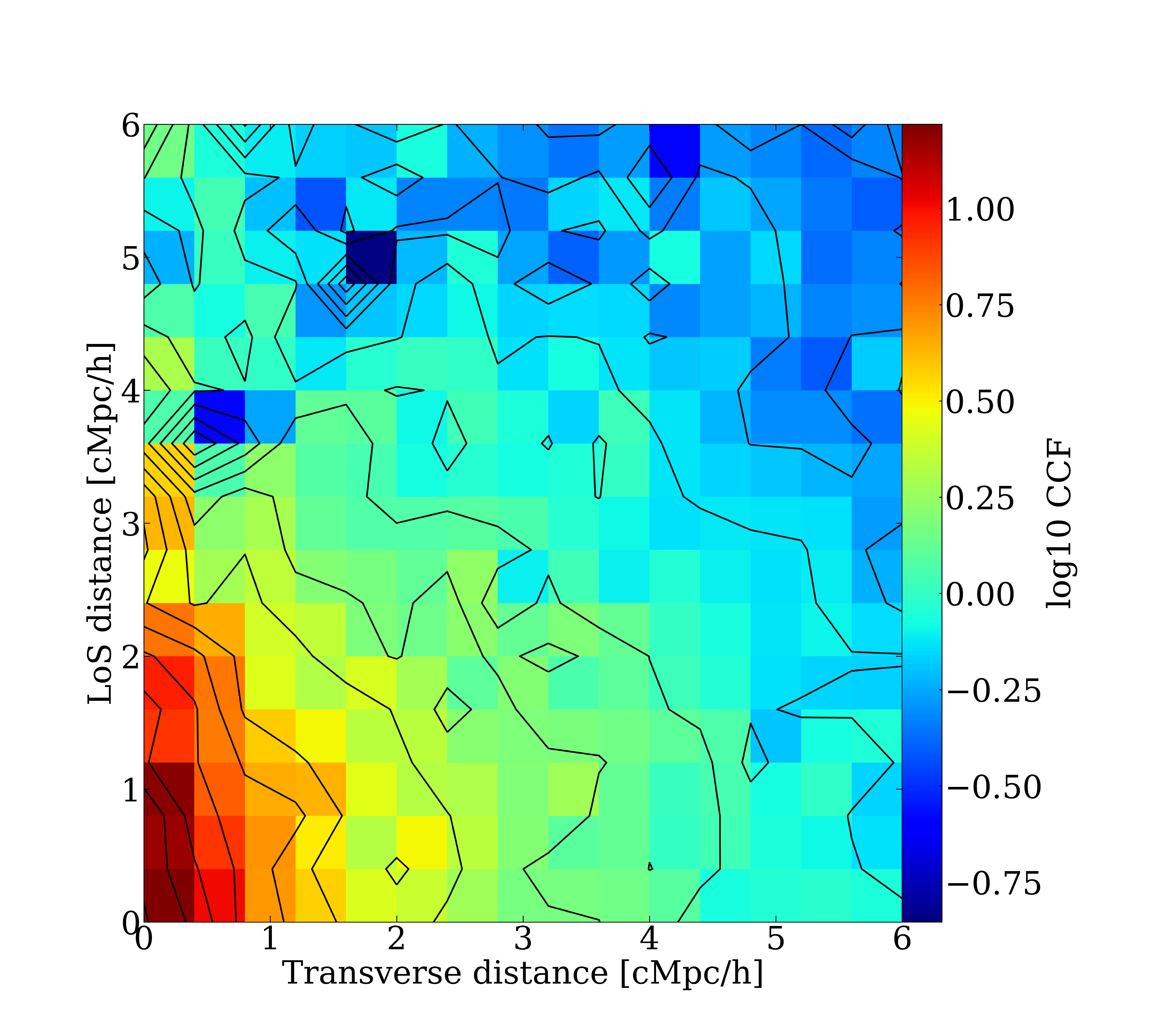}
    \caption{
   Two-dimensional CCF between galaxies and {\HI} with contours showing 16 levels between the min/max values ($-0.85 / 1.22$) in logarithmic scales. 
    }
\label{fig:ccf3D}
\end{figure}

Figure~\ref{fig:ccf3D} shows the 2D CCF between galaxies and {\sc Hi} with contours showing 16 levels between minimum and maximum on a logarithmic scale.  
Consistently with the previous observational and simulation results \citep[][]{Rakic12,Rakic13,Turner14,Turner17}, we find the finger-of-God signature of elongation along the LOS direction with a stronger CCF signal. 
We intend to carry out further analysis of gas dynamics (inflow/outflow) to figure out what's really causing this finger-of-God signature. 
To detect this effect, Figure~\ref{fig:ccf3D} shows that it requires a transverse resolution of $<1\,\hicmpc$. 
For example, \citet{Turner17} compared the results of the Keck Baryonic Structure Survey (KBSS) and the EAGLE cosmological simulations, and found good agreement in the 2D enhancement of {\HI} optical depth near galaxies out to 2\,pMpc, as well as detecting the finger-of-god-like feature. From the examination of  median ion mass-weighted radial velocities, they concluded that the feature was caused predominantly by the infalling gas motion rather than redshift errors. 

%%Fig. 9

%%%%%%%%%%%%%%%%%%%%%%%%%%%%%%%%%%%%%%%%%%%%%%%

\subsection{Comparison with the ``Mukae plot"}

% Fig. 10a,b
\begin{figure*}
\gridline{\fig{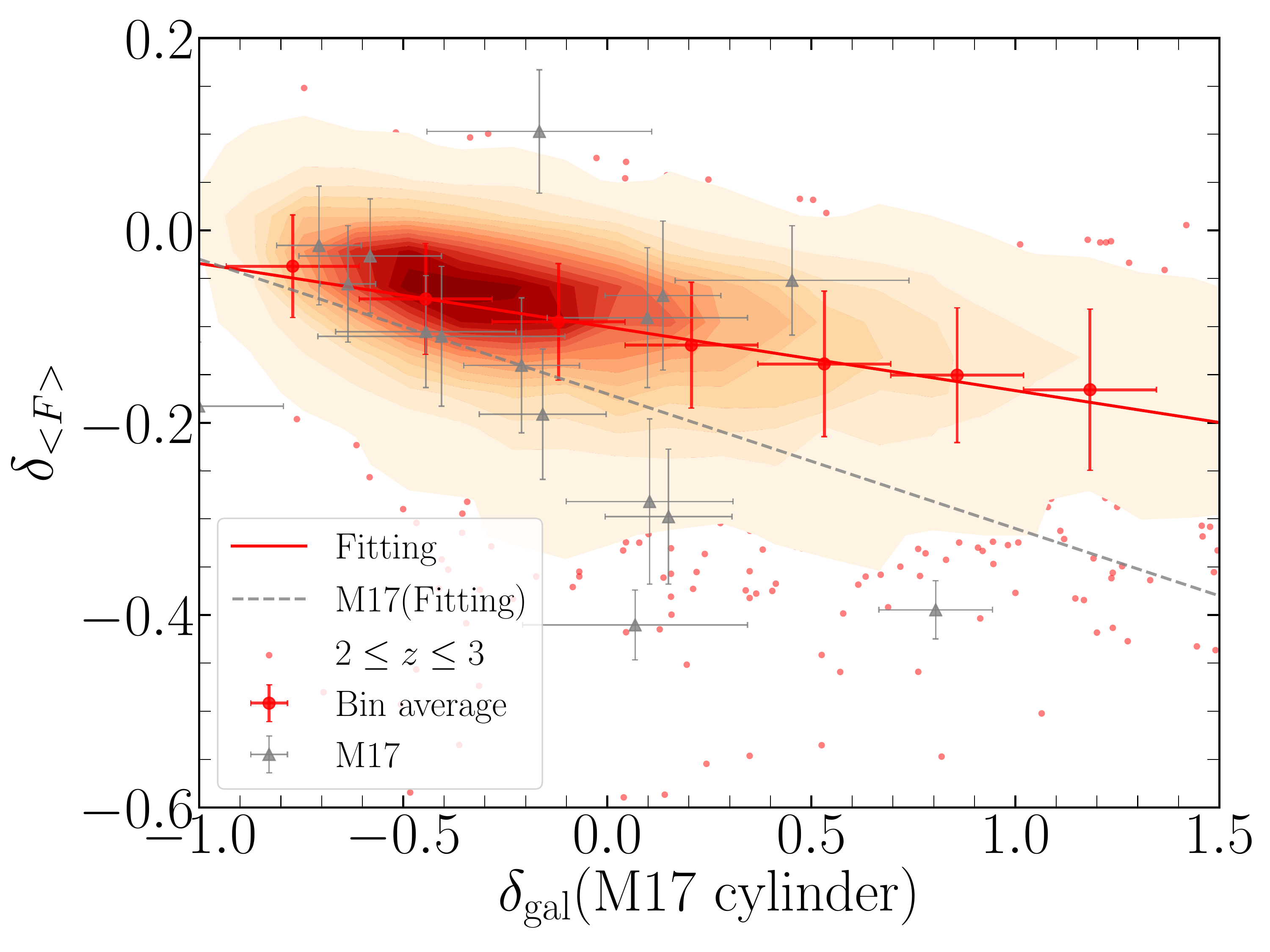}{0.5\textwidth}{(a)}
          \fig{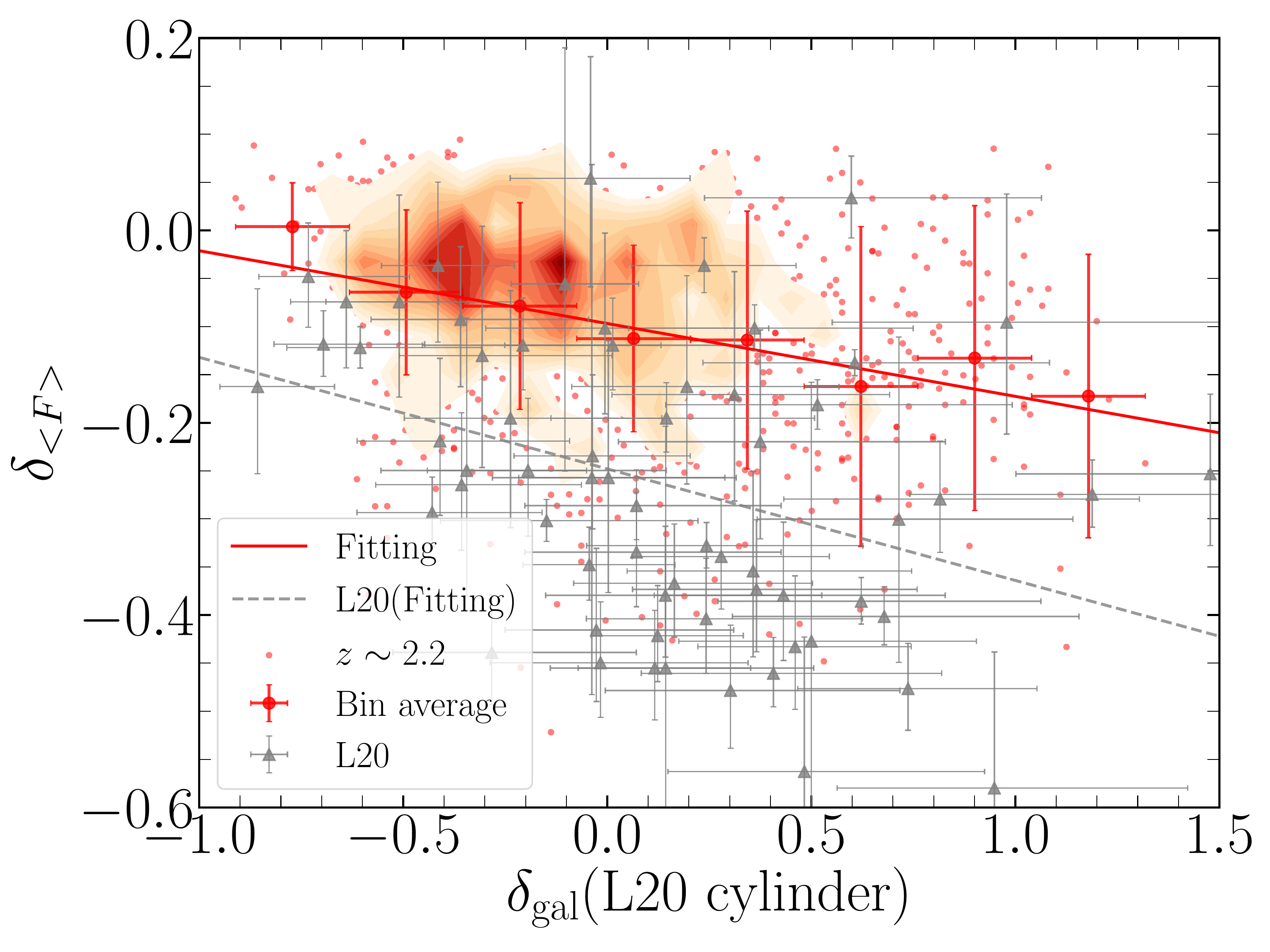}{0.5\textwidth}{(b)}}
    \caption{Relation between the flux contrast $\delta_F$ and galaxy overdensity $\delta_{\rm gal}$, compared against  observational data.
    {\it Panel (a)} compares the simulation result against the observational data of \citet[][gray data points and their linear regression fit shown by the grey line]{Mukae17}.  Each small red dot  corresponds to one cylinder along the LOS in our simulation, with their mean in each $\delgal$ bin (covering the entire distribution with eight equal-sized bins on the abscissa), and the linear regression fit to all the data points is shown by the red solid line.
     {\it Panel (b)} compares against \citet{Liang21}'s observational data. 
     Here, $\delta_F$ is computed within a length of $15\,\hicmpc$, i.e., $21.4$\,cMpc assuming $h=0.7$ along the LOS, and $\delgal$ is computed using a cylinder with an aperture radius ($10$\,cMpc) $\times 62.5$\,cMpc (length along the LOS $\approx$ FWHM of NB387 filter).  Note that the path length to compute $\delta_F$ and the cylinder size for $\delta_{\rm gal}$ are different in panels (a) and (b).  We use the red contour to avoid saturation of points; it is constructed with a $32\times 32$ grid with 16 equal logarithmic levels between minimum (four dots) and maximum surface density of red dots. 
     }
    \label{fig:Mukae}
\end{figure*}

Even if one does not have spectroscopic data, 
photo-$z$ data of galaxies are also useful to investigate macroscopic relations between gas and galaxies. The combination of spectroscopic and imaging surveys is powerful if we can  perform IGM tomography. For example, Figure 8 of \citet{Mukae20} compares the LAE overdensity filaments with gas filaments from the IGM tomography.

Using the photometric galaxy catalog from the COSMOS/UltraVISTA \citep{Muzzin13},
\citet{Mukae17} detected a weak correlation between galaxy overdensity ($\delgal$) and $\Lya$ absorption decrement ($\delta_F$) in cylinders with a base radius of $r=5^\prime$ ($\sim 2.5$\,pMpc at $z\sim 2.5$) and a cylinder length of 25 pMpc, which corresponds to their photometric redshift uncertainty. 
Their galaxies lie at $2<z<3$, and stellar masses were estimated by the SED fitting using the FAST code \citep{Kriek09}. They applied a criteria of $K<23.4$\,mag which corresponds to a stellar mass limit of $\log(\Mstar/\Msun) > 9.7$ at $z\sim 2.5$. 
In the following, we call this correlation plot between $\delgal$ and $\delta_F$ a ``Mukae plot." 
They found a weak negative correlation of 
\begin{equation}
\delta_F = \alpha\, \delgal + \beta
\end{equation}
with $\alpha =- 0.14^{+0.06}_{-0.16}$ and $\beta =-0.17^{+0.06}_{-0.06}$. 
 
In Figure~\ref{fig:Mukae}(a), we show the result from our light-cone data 
(with the total galaxy sample of $\Mstar>10^{9.7}\,\Msun$) estimated with the same cylinder size and $\Mstar$ limit as \citet{Mukae17} at $z=2-3$. 
The smaller red dots represent our measurements in each cylinder in our light-cone data, which show a greater scatter than Mukae's result.  The bigger red points with error bars are the average of smaller red dots, covering the entire distribution with eight equal-sized bins. 
The red solid line is the least-square fit to the median points with 1$\sigma$ error bars with 
$\alpha = -0.066 \pm 0.0011$ and $\beta = -0.100 \pm 0.00060$. %for M*>10^9.7 Msun sample
%$\alpha = -0.084 \pm 0.0021$ and $\beta = -0.099 \pm 0.00081$. %for M*>10^9 Msun sample 
Our simulation result has a shallower slope than the Mukae's fit. 
The number of data points of \citet{Mukae17} is much smaller than the number of our simulated data points, and their fit may have been affected strongly by the outliers with low $\delta_{\langle F \rangle}$ values ($\sim -0.4$). 
 
In Figure~\ref{fig:Mukae}(b), we compare our simulation result against the observational data by \citet{Liang21}, who made similar measurements of $\delgal$ (with 2642 $\Lya$ emitters at $z\sim 2.2$) using a cylinder with a base aperture radius of $10$\,cMpc and a length of $62.5$\,cMpc (the FWHM of the NB387 filter) along the LOS. 
As for $\delta_F$, the averaging along the LOS was done over $10\,\hicmpc$, i.e., $21.4$\,cMpc assuming $h=0.7$, and the center of an absorber was searched for as the absorption peak around $3,862\pm 35$\,\AA\, ($z\approx 2.15 - 2.20$).
They also found a positive correlation between LAE overdensity and the effective $\Lya$ optical depth estimated from 64 eBOSS quasar spectra.  
\citet{Liang21} gave the best-fit results of 
$\alpha = -0.116^{+0.018}_{-0.022}$ and $\beta = -0.248^{+0.082}_{-0.093}$ (for all four fields)
and $\alpha = -0.227^{+0.026}_{-0.023}$ and $\beta = -0.258^{+0.096}_{-0.114}$ (without the J0210 field).
 
Correspondingly, we made similar measurements using the total galaxy sample of  $\Mstar>10^9\,\Msun$ in our light-cone data with the same cylinder size and LOS path length, and the only difference of placing our cylinders at the redshift of concern, covering $\pm L/2$ where $L=21.4$\,cMpc.  
Our best-fit result gives
$\alpha = -0.0664\pm 0.00476$ and $\beta = -0.0938 \pm 0.00238$, which is close to Liang's result.  
Liang et al. used many more data points with more scatter than Mukae et al., which could be one of the reasons for the shallower slope. See their \S\,5.1 for further discussions on the sample bias. 
It is nevertheless interesting that a similar galaxy--{\HI} correlation is obtained even in an extreme system such as the MAMMOTH region \citep{Cai16,Liang21} to that of Mukae's more general field.  
The exact value of intercept parameter $\beta$ is probably not so important at this point, as each observational result is based on a different sample of galaxies with different normalization.

\citet{Liang21} also presented the results from our simulation using different galaxy samples of $\Mstar = 10^8-10^9, 10^9-10^{10},$ and $10^{10}-10^{11}\,\Msun$, and showed that the slope `$\alpha$' became shallower only by $\Delta \alpha \sim 0.3$ for the more massive galaxy sample. 
This trend is consistent with the naive expectation that the more massive galaxies are associated with deeper potential wells and more abundant {\HI} gas in them.  It is also consistent with our earlier results shown in Figure~\ref{fig:impact}(c) and (d), where a stronger flux contrast is detected around more massive galaxies. 

Given that the dependence of the slope on galaxy stellar mass is weak in our simulation, \citet{Liang21} proposed that the attenuation of $\Lya$ emission by the intergalactic {\HI} might be the more dominant effect in lowering $\delta_{\rm LAE}$ (and hence making the slope `$\alpha$' steeper) rather than the different galaxy stellar mass of the sample. 
The spatial offset of LAEs from general galaxies are also observed statistically by \citet{Momose21a,Momose21b}.
\citet{Mukae20} also pointed out the photoionization effect by the radiation from galaxies and quasars, producing proximity zones \citep[e.g.,][]{Bajtlik88,Scott00,Calverley11}. 
To verify these arguments, we would have to perform $\Lya$ radiation transfer calculations through the IGM, which we plan to carry out in the future. 

%%%%%%%%%%%%%%%%%%%%%%%%%%%%%%%%%%%%%%%%%%%%%%%

\subsection{Flux Contrast around a Protocluster}
\label{sec:proto}

% Fig. 11
\begin{figure*}
    \epsscale{1.33}
    \plotone{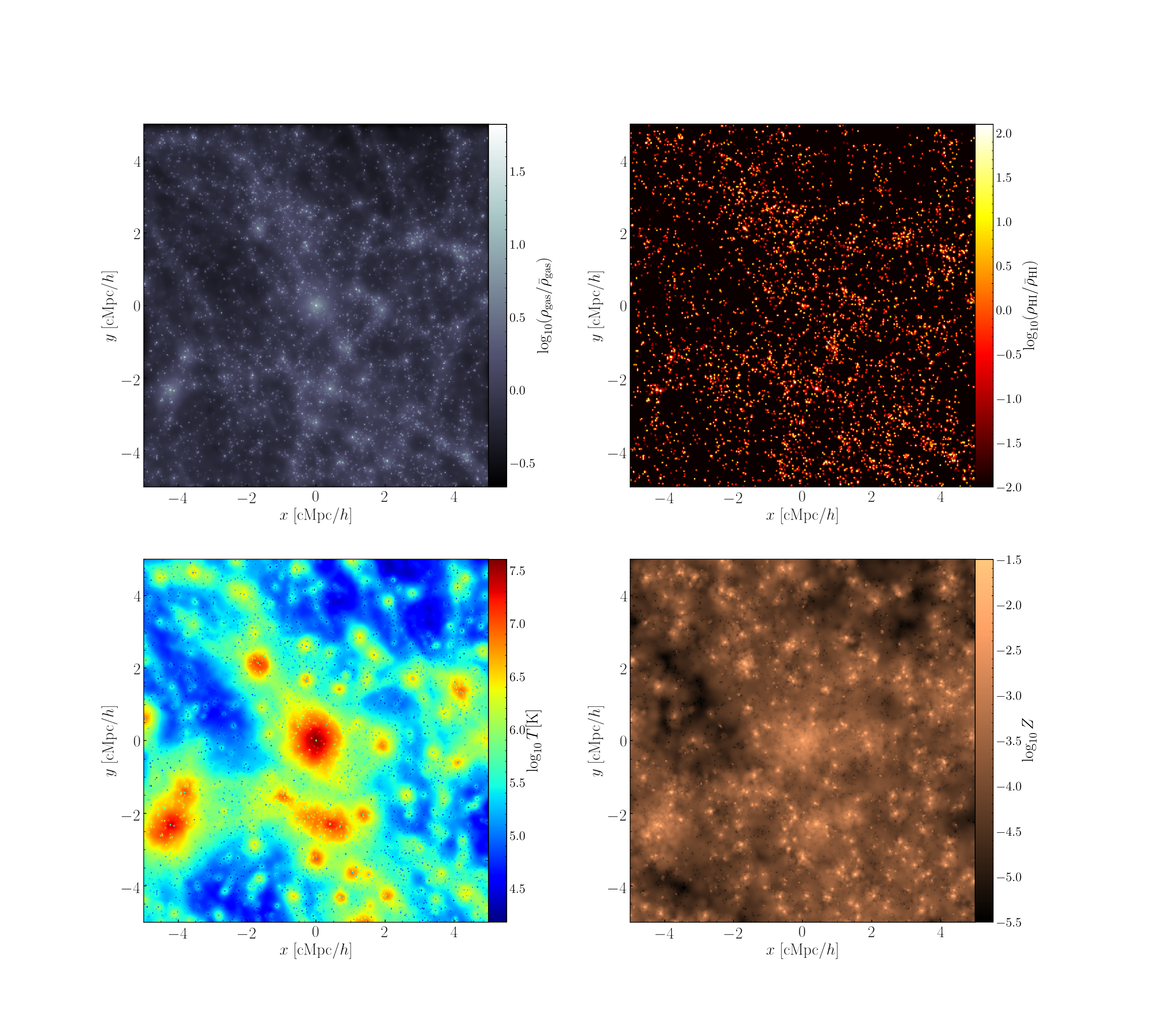}
    \caption{Projected maps of various quantities (gas overdensity, {\sc Hi} overdensity, temperature, and metallicity, from top-left to bottom-right, respectively) in the protocluster region, which is the most massive system in our light-cone data. 
    The mass of the protocluster is $M_{\rm h} = 9.4 \times 10^{13}\himsun$ at $z\sim 2.1$, and the virial radius is $R_{\rm v} = 1.1$\,cMpc.
  } 
    \label{fig:protocluster}
\end{figure*}
    
Protoclusters have been an interesting focus of high-$z$ galaxy studies. Since they are biased regions with early evolution of the density field, they can give us important insight into  early structure formation and help us to check the $\Lambda$CDM model during the critical phase of structure formation, when it is strongly driven by gravitational instability.  
Using protoclusters as cosmological probes can also highlight the importance of environmental effect of the cosmic star formation rate density \citep[e.g.,][]{Chiang17}, and the filamentary distribution of cosmic gas \citep[e.g.,][]{Umehata19}.

The definition of a protocluster may vary depending on the authors, but usually it either is a high concentration of galaxies at high redshift \citep[e.g.,][]{Toshikawa16,Toshikawa18} or has the specific meaning of progenitors of present-day galaxy clusters in numerical simulations.  In the actual observations, one cannot follow the time evolution of a system at a certain redshift to the present day, so one has to rely on other theoretical arguments, e.g., estimate the dark matter halo mass and refer to the results of $N$-body simulation for the statistical connection between high-$z$ and low-$z$ samples. 

Future IGM tomography studies will be able to identify numerous protoclusters without relying on spectroscopic follow-up of galaxies.  
In fact, \citet{Lee16a} discovered an extended IGM overdensity with deep absorption troughs at $z=2.45$ associated with a previously discovered protocluster at the same redshift.  They estimated that this IGM overdensity is associated with dark matter of mass $M_{\rm DM}(z=2.45) \sim 1.1\times 10^{14}\,\himsun$, which will later grow into a galaxy cluster with $M_{\rm DM}(z=0) \sim 3\times 10^{14}\,\himsun$. 
\citet{Krolewski18} also reported the discovery of a cosmic void at $z\sim 2.3$ using a 3D $\Lya$ tomographic map. 
Therefore it would be desirable to study  tomographic properties of protoclusters in cosmological hydrodynamic simulations with full physics, and see whether we find similar physical properties. 
 
Figure~\ref{fig:protocluster} shows the projected quantities around a protocluster at $z\sim 2.1$ with a halo mass of $M_{\rm h} = 9.4 \times 10^{13}\himsun$ and a virial radius of $R_{\rm v} = 1.1$\,cMpc.
This is the most massive system in our light-cone data set. 
There is a clear high-overdensity region in the center of the upper left panel of the gas density field. Correspondingly the high-temperature ($T\gtrsim 10^7$\,K) intracluster gas is also visible in the center of the lower left panel; this would emit X-rays. 
The central region of this protocluster is already enriched to $\log(Z/Z_\odot) \sim -1.5$ by $z=2.1$. 
On the other hand, it is interesting to see that the protocluster is not so visible in the {\HI} overdensity map in the upper right panel, because the {\sc Hi} gas is more localized in the centers of galaxies and not as prominent as the hot X-ray-emitting intracluster gas within the protocluster. 

\begin{figure}
    \epsscale{1.2}
    \plotone{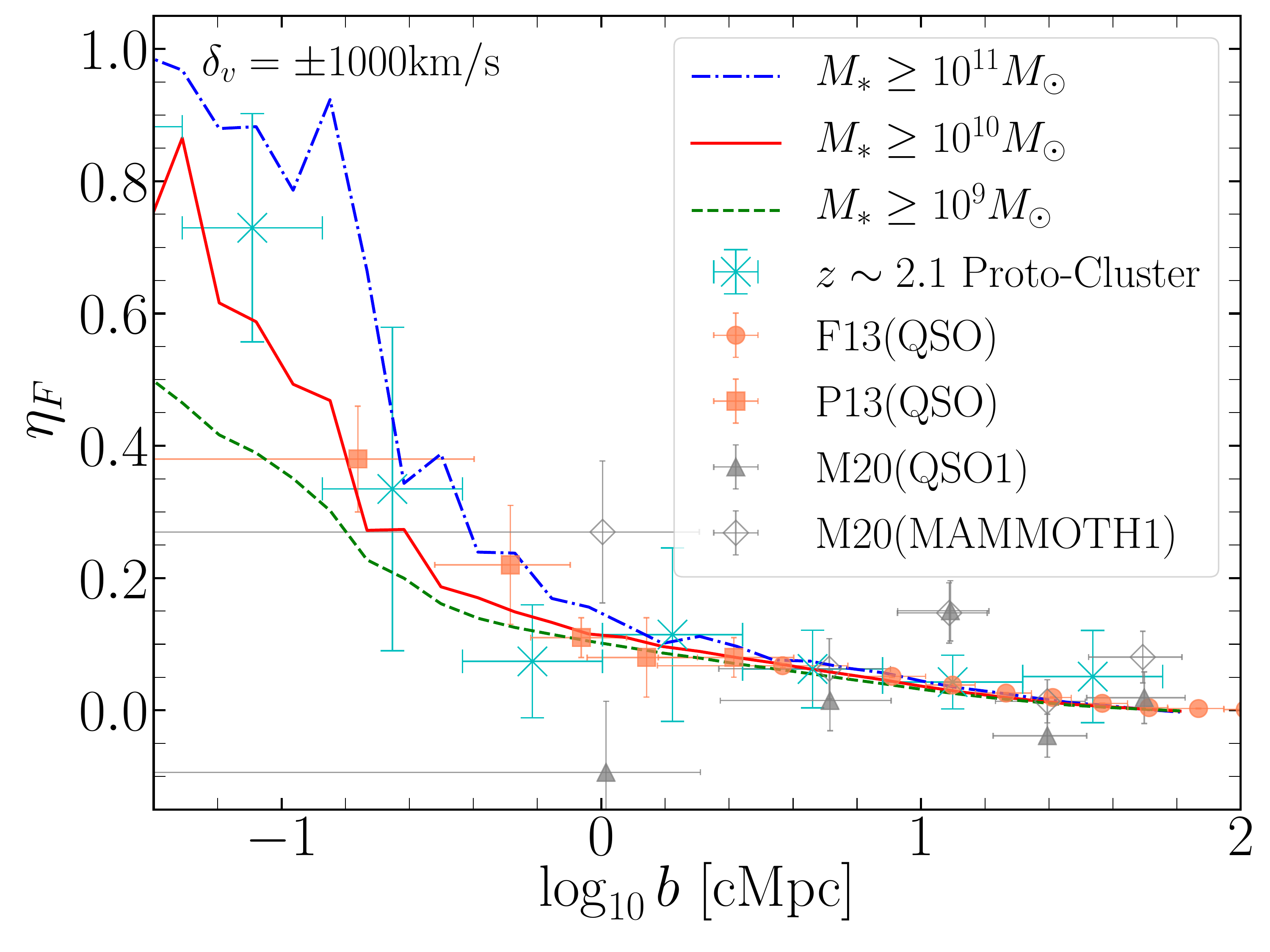}
    \caption{Flux contrast around the protocluster shown in Figure~\ref{fig:protocluster} at $z\sim 2.1$ as a function of transverse distance from the protocluster center.  The cyan crosses show the average flux contrast in each transverse bin, and each data point is an average of eight LOSs with the error bar showing the min/max scatter.  The other observational data points and lines are the same as in Figure~\ref{fig:impact}(d) and shown for comparison. Additionally, black open circles are from \citet{Mukae20}, possibly indicating the quasar proximity effect. } 
    \label{fig:PCflux}
\end{figure}

Figure~\ref{fig:PCflux} shows the flux contrast as a function of transverse distance from the center of the protocluster at $z\sim 2.1$ as cyan crosses, and it covers the range of $b=[0.02, 64]$\,cMpc with eight equal-sized logarithmic bins.
Note that the observational data points are shown only for comparison purposes, and we caution that each data set probes a different median redshift from the protocluster. 
On large scales of $b>1$\,cMpc, the flux contrast signal converges to the large-scale IGM signal, but we see an enhancement of signal at $b<200$\,ckpc toward the center of the protocluster, even though it is not really visible to our eyes in the upper right panel of Figure~\ref{fig:protocluster}.  
In particular, in the innermost bin of $b\simeq 100$\,ckpc, we see an enhancement by a factor of $\sim 2$ over the observed data points. 
The protocluster flux contrast lies between those for the galaxy population of $\Mstar \ge 10^{10}$ and $\Mstar \ge 10^{11} \Msun$. 
This is reasonable, because we are approaching to the protocluster center where the brightest cD galaxy may be developing in the core. 
Our result is also in line with that of \citet{Miller19}, in the sense that the high-$\tau$ sightlines are dominated by galaxies and DLAs, whereas most of the sightlines simply have lower $\tau$. When the contributions from protocluster galaxies are averaged over in spherical shells, this can produce the increasing flux contrast toward the protocluster center. 

% Fig. 12

In Figure~\ref{fig:PCflux}, we also show the data points from \citet{Mukae20}, who studied the 3D distribution of {\HI} around quasars in the MAMMOTH-1 nebula \citep{Cai17} at $z=2.3$ using the {\HI} tomography technique. 
Their 3D radial {\HI} radial profile showed a weaker absorption near the MAMMOTH-1 QSO, suggesting a proximity zone of photoionized gas \citep[e.g.,][]{Bajtlik88,Scott00,Calverley11}.
Here, for a fair comparison with our simulation results, we show the ``R2D" estimate taken from their Figure\,A1 in the Appendix for MAMMOTH1-QSO and QSO1. 
The MAMMOTH1-QSO result is closer to the P13 data, but the QSO1 data show a stronger decline in $\eta_F$ on small scales of $\log (b[{\rm cMpc}])\simeq 0.0$.  

We also note that \citet{Momose21b} detected weaker {\HI} absorption within the central few cMpc around AGNs, and that the maximum absorption is at $r=5-6\,\hicmpc$, again suggesting a proximity effect. 
The differences between the data points of \citet{Pro13} and \citet{Mukae20}'s QSO1 data can be reconciled with anisotropic emission from quasars (see Section~6.3 of \citet{Pro13} for further discussions).
Therefore, if we want to study the 3D distribution of {\HI} around quasars and AGNs, a realistic model of AGN feedback is desired in simulations. 
On the other hand, \citet{Sorini20} showed that, on average, the impact of AGNs is less strong than that of stellar feedback on $\Lya$ forest statistics.  
It is clear that more detailed studies are needed to figure out when and how AGN feedback affects $\Lya$ forest statistics, and we plan to pursue this in the future.

%%%%%%%%%%%%%%%%%%%%%%%%%%%%%%%%%%%%%%%%%%%%%%%%%%%%%%%%%%%

\section{Discussion \& Summary}
\label{sec:conclusion}

In preparation for the upcoming Subaru PFS survey and other major telescopes such as TMT and JWST, we produce the light-cone and LOS data sets at $z=2-3$ using the {\sc GADGET3-Osaka} cosmological hydrodynamic simulations with star formation and SN feedback models. 
We compute various $\Lya$ forest statistics including 1D flux PDF, 1D power spectrum, cross-correlation function, and flux contrast as a function of transverse distance from galaxy centers and a protocluster. 

We draw 1024 LOSs to compute the $\Lya$ forest data, and find that the 1D flux PDF shows 
a rough agreement with observations, although with some deviations depending on the value of $F$ and redshift. Our jackknife estimate of 1$\sigma$ error is comparable to that of the observational estimates, but the sample variance is larger than the jackknife estimate, consistently with previous work \citep[][]{Kim04,Rollinde13}. 
We checked that the variance does not change very much even if we draw 16,384 LOSs, which means that our results do not change very much by increasing the number of LOS further.   
The redshift evolution of the flux PDF is as expected, where the fraction of low-transmission lines increases with increasing $\tau_{\rm eff}$ toward higher redshift.  
We find that the different models of SN feedback and UVB self-shielding treatment causes a variance of flux PDF up to $5-10$\% depending on the model, which is comparable to the jackknife errors, and so not too significant.

The simulated 1D $\Lya$ power spectra show more intricate deviations from observations, and require further analyses to understand them. 
Our Fiducial model captures the higher $k$ modes well at $\log (k\, {\rm [s/km]}) > -1.3$ at $z\sim 2.4$, but at the same time underpredicts the power at intermediate scales of $\log (k\, {\rm [s/km]}) \sim -1.5$, and then overpredicts at the large scales of $\log (k\, {\rm [s/km]}) < -2.5$. This may have to do with the fact that we have not tuned our thermodynamic parameters (e.g., gas temperature) assuming the FGPA method. 
The results of the NoFB and CW runs deviate from the Fiducial run only within the 1$\sigma$ jackknife error, but the Shield and FG09 runs deviate from the Fiducial run by more than 1$\sigma$. The deviation is the greatest for the FG09, probably due to the difference in the overall photoionization rate from that of the HM12 model adopted in the Fiducial run as we discussed in Section\,\ref{sec:Pk}.
In the future, we will further examine the 1D $\Lya$ power spectra using simulations with a higher resolution and larger box size. 

Perhaps the more direct statistic is the flux contrast $\eta_F$ as a function of impact parameter $b$ from galaxies (Figure~\ref{fig:impact}). On large scales of $b > 1\,\hicmpc$, we find good agreement with observational estimates and the previous work by \citet{Sorini18}, which supports the correctness of the $\Lambda$CDM model and the large-scale structure that it predicts.  
On the other hand, on small scales, we find variations due to different parameters, such as the galaxy mass, probing depth, redshift, and feedback models. 
Overall our results show good agreement with currently available observations; however, we find an important dependence of $\eta_F$ on galaxy stellar masses, where more massive galaxies are surrounded by more abundant {\HI} gas with higher flux contrast.  If one combines all the signals from all galaxies and averages over them, then the stronger signal from more massive galaxies ($\Mstar\ge 10^{11}\,\Msun$) is diluted by that of lower-mass galaxies, and the flux contrast appears to be reduced in the proximity of galaxies. 

We also see some ($\sim$20--30\%) variations in $\eta_F$ on small scales depending on the physical treatment of feedback, self-shielding of gas, star formation, and UVB (Figure~\ref{fig:impact}(e)).  It is somewhat counterintuitive that the NoFB run has the lowest contrast, which is probably because much of the cold gas is consumed by star formation in the dense galactic center. 
The CW run is the second lowest, owing to its efficient removal of gas from the galactic potential with strong galactic wind. The Fiducial and the Shield models have about the same level of $\eta_F$ on small scales.  The FG09 run has the highest level of $\eta_F$ relative to the HM12 model in the Fiducial run, perhaps due to its lower photoionization rate (see Section~\ref{sec:Pk}).  
While these deviations are interesting to discuss, the variation due to galaxy stellar mass and the probed path length ($\delta_v$) is larger than that due to different feedback models.  The redshift evolution of $\eta_F$ is clearly seen, and the contrast at $z\sim 3$ is higher than that at $z\sim 2$ at all scales of $100\,{\rm ckpc} < b < 20\,{\rm cMpc}$ due to higher $\tau_{\rm eff}$. 
At the same time, we caution the reader that the results on $\eta_F$ may not be fully converged due to its strong dependence on the box size and galaxy sample (see Appendix~\ref{app:res}). 

We find a strong flux contrast around a protocluster at $z\sim 2$ in our simulation, which is comparable to that for the galaxies with $\Mstar \ge 10^{10}-10^{11}\,\Msun$ (Figure~\ref{fig:PCflux}).  
This result certainly lends support to finding more protoclusters via {\sc Hi} tomography and massive galaxies associated with it. 

The cross-correlation between galaxies and {\sc Hi} absorbers shows expected trends. For example, the higher-EW sample is more correlated with galaxies than the lower-EW sample, because more neutral {\sc Hi} gas is expected to be closer to the galaxies.  
More massive galaxies have stronger CCF signal on larger scales of $>500$\,ckpc, but the signal becomes noisy at smaller scales due to lack of a correlating sample.  
The redshift evolution of the CCF signal is not so strong, but we do see that it is stronger at $z=2$ than at $z=3$ at least on larger scales of $>1$\,cMpc, as the cosmic structure develops and galaxy clustering becomes stronger. 
Our 2D CCF result (Figure~\ref{fig:ccf3D}) shows the finger-of-God feature similarly to \citet{Turner17}, but it would require a transverse resolution of $<1$\,cMpc resolution in the IGM tomography to study this effect in more detail observationally. 
Using our simulations, we intend to study the gas dynamics that causes these effects in more detail in our subsequent work, e.g., the impact of cold accretion flows and galactic outflows on the 2D CCF. 

We also examined the correlation between galaxy overdensity $\delgal$ and  flux contrast $\delta_F$, and compared the results with those of \citet{Mukae17} and \citet{Liang21} using the same cylinder size as their measurements.  Our simulation result shows a greater scatter due to larger  sample size than in the observations, and a shallower best-fit slope than the observations (Figure~\ref{fig:Mukae}). 
These differences could be due to a few things, such as the different galaxy samples used, cosmic variance, or the effect of photoionization by the radiation emitted by galaxies or quasars. 
This comparison study using the Mukae plot is not as definitive as the flux contrast at this point; however, it provides a support for such a methodology utilizing photo-$z$ galaxy samples that are more widely available with lower observational costs. 

The examination of a protocluster in our simulation (Figures~\ref{fig:protocluster}, \ref{fig:PCflux}) also brings up interesting directions for our future research. In particular, we see a hot X-ray-emitting gas in the center of the protocluster, but at the same time we do not see a diffuse {\HI} distribution near the center.
Rather, it seems that the {\HI} absorption is mostly contributed by individual galaxies rather than a diffuse intra-protocluster medium, and when averaged over the spherical shells, it produces an enhancement of flux contrast in the central few hundred ckpc. 
This result seems to be in line with that of \citet{Miller19}, in the sense that the high-$\tau$ sightlines are dominated by galaxies and DLAs, while most of the sightlines simply have lower $\tau$. 
In our future work, we plan to follow up on these issues utilizing higher-resolution zoom-in simulations of protoclusters. 

In this paper, we focused on the relative distribution of galaxies and {\HI}, and did not touch on the metal absorption lines, which we will discuss in our future publications.  Cross-correlation studies between galaxies, {\HI}, and metals will certainly give us useful information on the interplay between them and enable us to learn more about star formation and feedback. 
The astrophysics in this subject is quite rich, and at the same time quite challenging to get it all correctly, because we need to have a good understanding of cosmology, galaxy formation, star formation, feedback, and chemical enrichment of the IGM, and link them all together to have a full picture.

\acknowledgments
We thank the anonymous referee for detailed comments that helped to improve the manuscript significantly. 
K.N. acknowledges useful discussions with J.X. Prochaska, and is grateful to V. Springel for providing the original version of {\sc GADGET-3}, on which the {\sc GADGET3-Osaka} code is based. 
We also thank Yuri Oku for providing the L40N512 run for the comparison presented in the appendix. 
Our numerical simulations and analyses were carried out on the XC50 systems at the Center for Computational Astrophysics (CfCA) of the National Astronomical Observatory of Japan (NAOJ), {\sc Octopus} at the Cybermedia Center, Osaka University, and {\small Oakforest-PACS} at the University of Tokyo as part of the HPCI system Research Project (hp180063, hp190050, hp200041).
This work is supported by the JSPS KAKENHI grant No. JP17H01111 (K.N.), 19H05810 (K.N.), 20H00180 (K.N., M.O.), JP18H05868 (K.G.L.), and JP19K14755 (K.G.L.).
K.N. acknowledges the travel support from the Kavli IPMU, World Premier Research Center Initiative (WPI), where part of this work was conducted. 
The authors thank the developers of below software for making their code available on a free and open-source basis.

%% Similar to \facility{}, there is the optional \software command to allow authors a place to specify which programs were used during the creation of the manusscript. Authors should list each code and include either a citation or url to the code inside ()s when available.

\software{  CELib \citep{Saitoh17},
            FFTW (\url{http://www.fftw.org/}),
            Grackle \citep{Smith17},
            GSL \citep[][\url{https://www.gnu.org/software/gsl/}]{GNU},
            HDF5 (\url{https://portal.hdfgroup.org/}),
            Matplotlib \citep{Hunter07},
            Numpy \citep{NumPy20},
            SciPy \citep{SciPy20}
            %astropy \citep{2013A&A...558A..33A},
%          Cloudy \citep{2013RMxAA..49..137F}, 
%          SExtractor \citep{1996A&AS..117..393B}
        }

%%%%%%%%%%%%%%%%%%%%%%%%%%%%%%%%%%%%%%%%%%

\appendix

\section{Observational Information} 
\label{app:obs}

In the 2020s, we enter the era of massively multiplexed spectroscopy,  which enables us to study the IGM from small ($\sim$\,kpc) to large scales ($\sim$\,100\,cMpc).  This means that a unique era is coming soon where we can test the simulated universe by observations with a greater precision (i.e., the era of `{\it precision structure formation}').  In this appendix, to set the stage, we review some of the details of Subaru PFS and other near-future facilities. In particular, there is at the time of writing no published description of the planned PFS IGM tomography program, and so we will provide some details here.

\subsection{Subaru PFS (2023)}
The Prime Focus Spectrograph \citep[PFS;][]{Sugai15} is expected to be installed on the prime focus of Subaru Telescope in 2021--2022.  Its wide field of view (FOV, 1.25 deg$^{2}$) and large collecting area (8.2\,m in diameter) are unique among other facilities.  There is continuous wavelength coverage of 3800\,\AA\ --  1.26\,$\mu$m with three cameras, at medium resolving power ($R\equiv \lambda/\Delta \lambda \sim$2000--4000).  
Currently, a large collaboration is planning to execute a survey on $\sim$300 nights from 2023 through 2027, called the Subaru Strategic Program (SSP). This SSP is envisaged as having three components: the 
Cosmology Survey, the Galactic Archeology Survey, and the Galaxy Evolution Survey --- an early version of this SSP plan is outlined in \citet{Takada14}, although the details therein are no longer up-to-date. 

The PFS Galaxy Evolution Survey, in particular, will be a deep survey covering approximately 14 deg$^2$ across three of the Subaru Hyper Suprime Camera Deep imaging fields \citep{Aihara18}.
Across multiple visits, over 340,000 spectra will be obtained across multiple target classes in the range $0.7\lesssim z \lesssim 7$.
One of the major goals of this project is to generate a 3D tomographic map of the optically thin \HI\ at $2.1<z<2.6$ with average transverse sightline separation of $\sim 4$\,cMpc using 16,000 star-forming galaxies ($g<24.7$; see more details in Table~\ref{tab:pfs}) at $2.5< z < 3.5$ acting as background probes of intervening absorbing neutral gas (i.e. IGM tomography). While this sightline density is similar to that of the CLAMATO Survey \citep{Lee18}, the PFS program will be approximately 50$\times$ larger in area and volume, covering a combined
comoving volume of $4\times 10^7 \mathrm{cMpc}^3$.  One of the primary products
envisaged from these observations is Wiener-filtered absorption maps of the large-scale structure
traced by the diffuse Ly$\alpha$ forest using the sample described in Table~2, which will provide sufficient detail to recover the cosmic web on scales of $\sim 4\,\himpc$ \citep{Lee16b, Krolewski17}. More recently, \citet{Horowitz19} presented the TARDIS analysis framework, which allows direct inference of the underlying density field, which allows reconstruction of smaller-scale structures ($\sim 2\,\himpc$) than Wiener filtering. 

\begin{deluxetable*}{lclccc}[h]
\label{tab:pfs}
\tablecaption{Subaru PFS IGM Tomography Targets}
\tablehead{
\colhead{Target Class}&
\colhead{Redshift }&
\colhead{Selection} &
\colhead{Exposure }&
\colhead{Targeted Objects}&
\colhead{Number/PFS FOV} \\
  & \colhead{Range} & & \colhead{Time (hr)} & \colhead{(Useful Spectra)} & \colhead{(1.25 deg$^2$)}
}
\startdata
IGM background (bright) & 2.5-3.5 & $y<24.3$, $g<24.2$ & 6 & 8300 (5810) & 690 \\
IGM background (faint) & 2.5-3.5 & $y<24.3$, $24.2<g<24.7$ & 12 & 14,000 (9800) & 1170 \\
IGM foreground & 2.1-2.6 & $y<24.3$ & 6 & 22,000 (15,400) & 1830 \enddata
\end{deluxetable*}

In addition to the LBGs and QSOs targeted as Ly$\alpha$ forest background sources, 
there will also be a category of ``IGM foreground" galaxies selected to be at $2.1 < z < 2.6$, i.e., coeval with the Ly$\alpha$ absorption. This class of targets will be broadly selected among galaxies with $y< 24.3$ (see Table~2) to span a wide range of galaxy properties.
Subaru PFS spectra should have sufficient spectral resolution and signal-to-noise ratio to measure the IGM metal absorption lines (e.g., Mg\,{\sc ii}, Si\,{\sc iv} and C{\sc iv}) simultaneously, as well as O{\sc ii} emission in the near-infrared. With approximately 15,000 useful spectra expected from this target class, we expect to be able to robustly test the trends discussed in this paper. \\

%%%%%%%%%%%%%%%%%%%%%%%%%%%%%%%%%%%%%%%%%%%%%%%%%%%%%%

\section{Box size and Resolution test of simulations} 
\label{app:res}

\begin{figure}[htb]
    \epsscale{1.2}
    \plotone{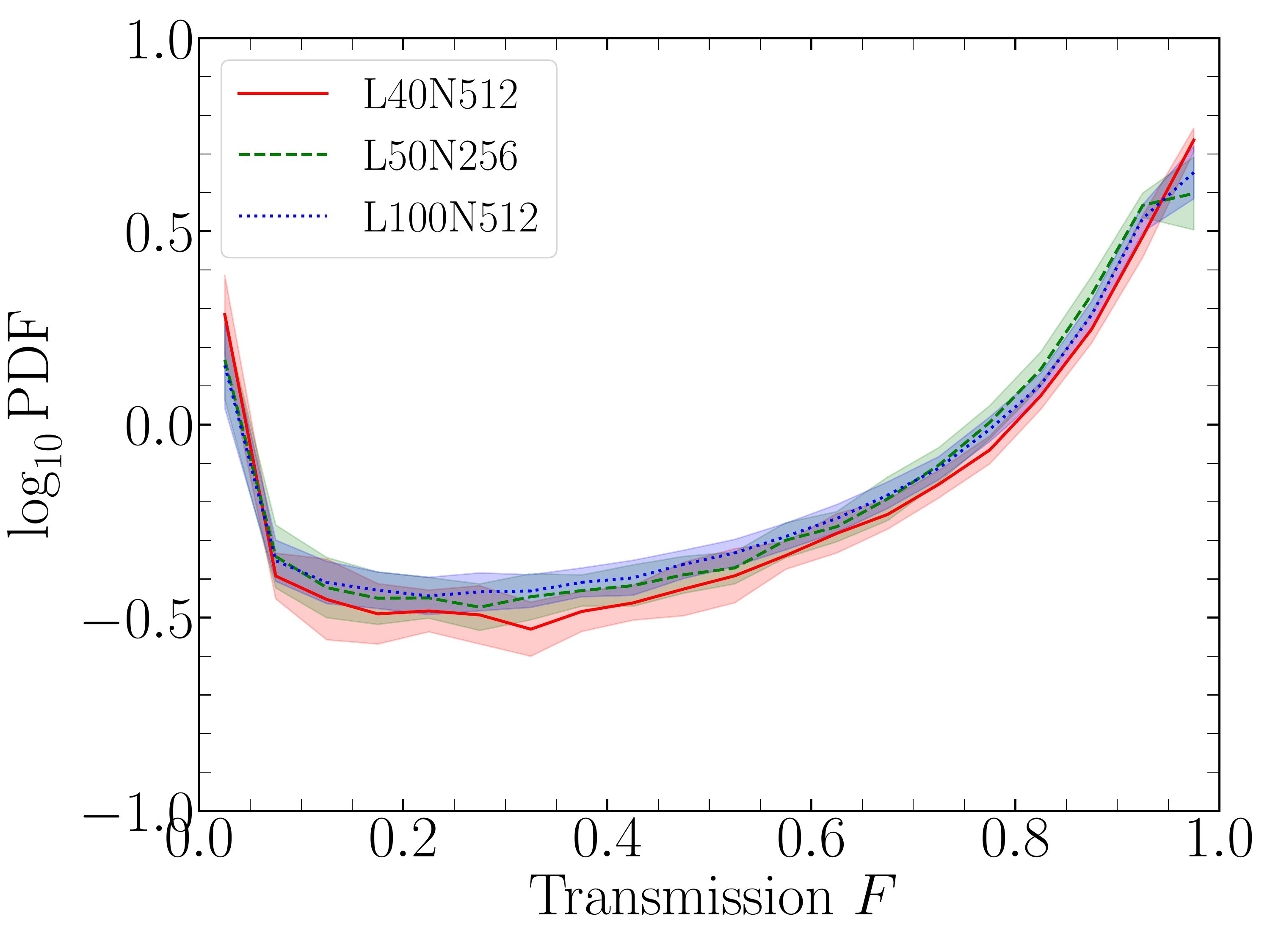}
    \caption{Flux PDF from the three simulations with different resolution and box sizes at $z=3$ (using a single snapshot of the Fiducial model; see Table~\ref{tab:res}). The same number of bins as in Figures~\ref{fig:fluxPDF}(c)--(e) is used. 
    Two simulations, L100N512 and L50N256, have the same mass and spatial resolution but with different box sizes, while the L40N512 has the highest resolution. See Table~\ref{tab:res}. 
    } 
    \label{fig:PDFcomp}
\end{figure}

In this appendix, we examine the effects of different box sizes and resolutions for the Fiducial model on the basic statistics presented in this paper. The two simulations listed in Table~\ref{tab:res}, L50N256 and L100N512, have the same mass and spatial resolution but  different box sizes, while the L40N512 has a higher resolution.
In Figure~\ref{fig:PDFcomp} we compare the flux PDFs from three simulations at $z=3$ (just using a single snapshot of each simulation). The results of L50N256 and L100N512 overlap well within 1$\sigma$, but L40N512 lies below the other two simulations at $0.1<F<0.9$ by about 1$\sigma$ and higher at the lowest and highest $F$ bins instead. Due to the higher resolution, the L40N512 run may have more optically thick lines with $F<0.1$, which instead pushes the PDF downward at $F>0.1$. Note that the ordinate is on a log-scale, therefore the differences at $0.2<F<0.8$ are much smaller than those at $F<0.1$ and $F>0.9$. 

\begin{figure}[htb]
    \epsscale{1.2}
    \plotone{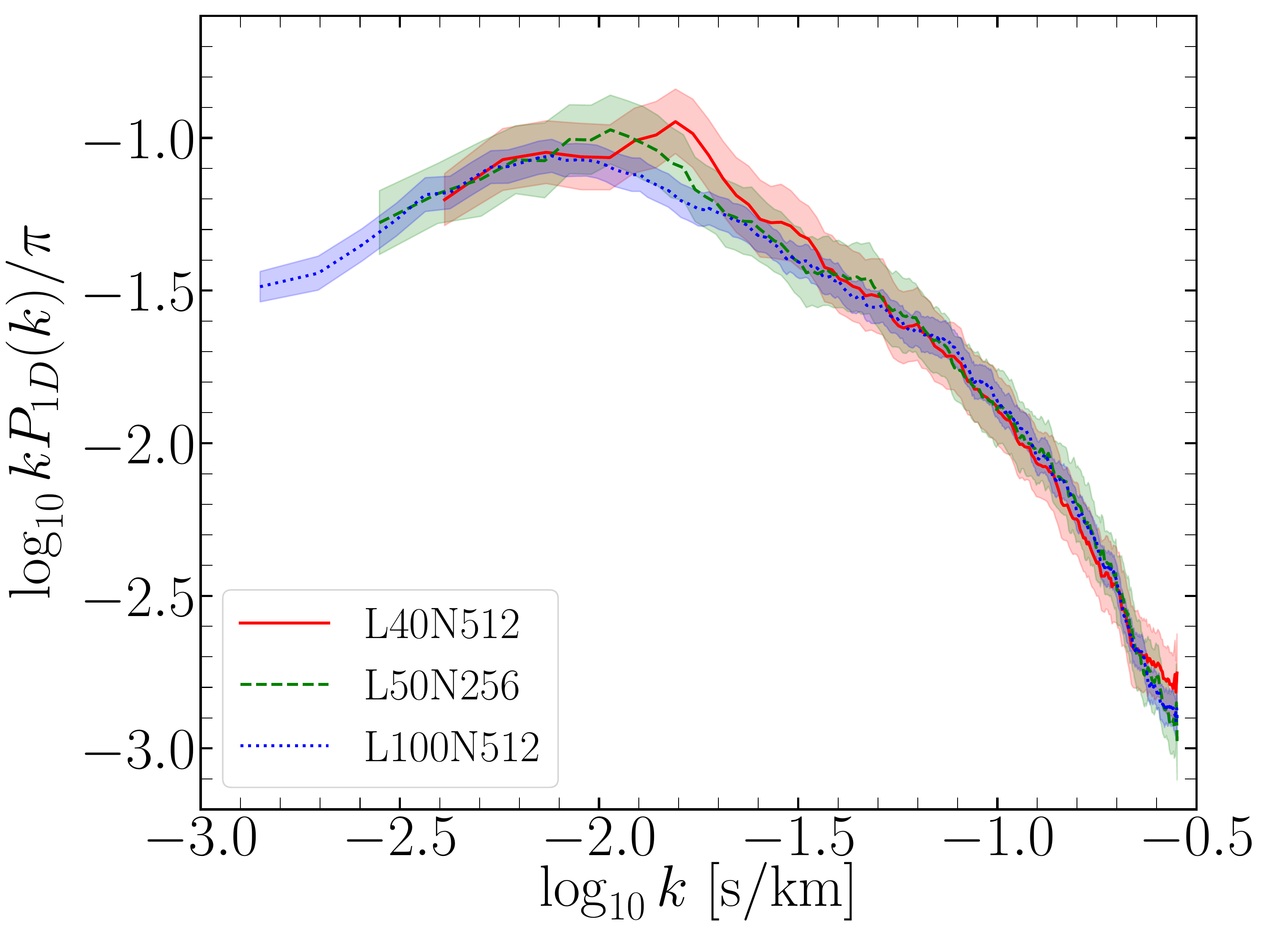}
    \caption{One-dimensional power spectra from the three simulations with different resolutions and box sizes at $z=3$ (see Table~\ref{tab:res}).  
    } 
    \label{fig:Pkcomp}
\end{figure}

In Figure~\ref{fig:Pkcomp}, we compare the $\Pk$ from the three simulations listed in Table~\ref{tab:res}. 
The three lines overlap well within the 1$\sigma$ jackknife error at $\log (k\,[{\rm s/km}])>-1.5$. There is a bump in the intermediate range for L40N512 and L50N256 at $\log (k\,[{\rm s/km}]) \sim -1.3$ and $-1.95$, which could be due to the limited simulation volume used here for the test. 

\begin{figure}[htb]
    \epsscale{1.2}
    \plotone{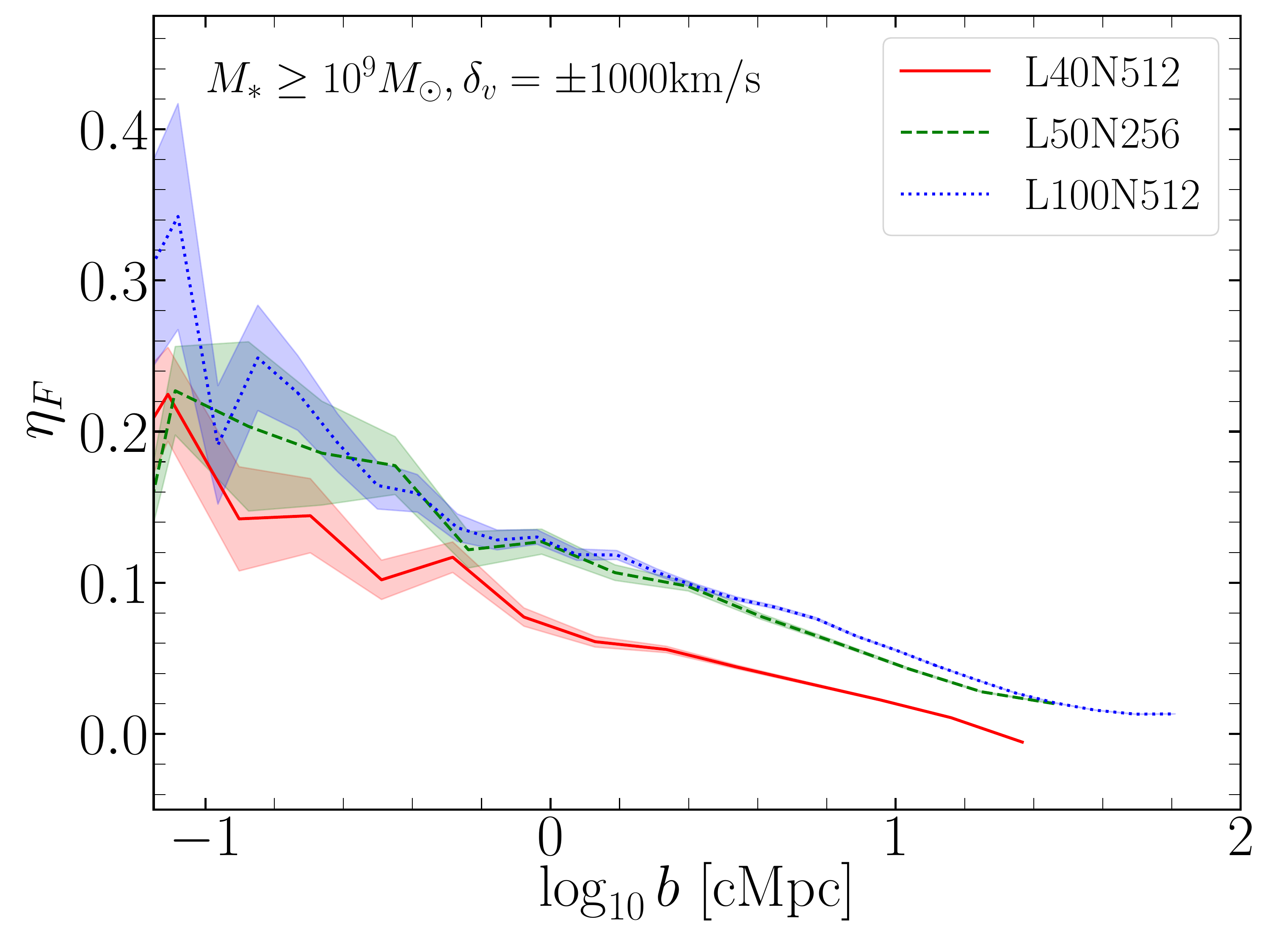}
    \caption{Flux contrast from the three simulations with different resolutions and box sizes at $z=3$ (see Table~\ref{tab:res}).  
    } 
    \label{fig:etacomp}
\end{figure}

Figure~\ref{fig:etacomp} compares the flux contrast $\eta_F$ for the same three simulations shown in Figures~\ref{fig:PDFcomp} and \ref{fig:Pkcomp}.
The results of L50N256 and L100N512 roughly agree within 1$\sigma$ at $-1\lesssim \log (b\,[{\rm cMpc}]) \lesssim 0.5$, but they deviate somewhat from each other at the smallest bin of $\log (b\,[{\rm cMpc}])<-1.0$ and on larger scales of $\log (b\,[{\rm cMpc}])\gtrsim 0.5$. As the box size becomes smaller from L100 to L40, $\eta_F$ converges to zero at a smaller impact parameter, near the half of the  simulation box size. The $\eta_F$ of L40N512 is lower than that of the other two runs (except at the smallest bins of $\log (b\,[{\rm cMpc}]) \lesssim -1$), which cannot simply be explained by the box-size effect.  With a higher resolution, L40 has a  higher star formation rate density than L50 and L100 runs at all redshifts by about a factor of 2, which leads to more consumption of {\HI} gas into stars and overproduction of lower-mass galaxies in lower-mass halos (because we did not recalibrate the feedback parameters as the resolution was increased). This could result in less {\HI} gas around the galaxies. In other words, the difference in resolution induces differences in the star formation rate and feedback \citep[e.g.,][]{Meiksin14,Meiksin15}, so the convergence test requires further detailed analyses on the star formation rates and galaxy populations, which we plan to present in the future.  
The $\eta_F$ of L40N512 and L50N256 agree at the smallest scale, but the L100N512 result is higher than the former two runs in the innermost bin. This is because Figure~\ref{fig:etacomp} was produced using only a single snapshot and the volume is limited, with a limited number of massive galaxies with $\Mstar > 10^{10}\,\Msun$. Therefore the L100N512 run with the largest number of massive galaxies (see Table~4) is exhibiting the strongest $\eta_F$ signal on small scales, here again showing the strong dependence of $\eta_F$ on the galaxy stellar mass as we argued in Section~\ref{sec:contrast} and Figure~\ref{fig:impact}(c),(d).
Therefore we caution the reader that the results of $\eta_F$ could depend strongly on the box size and associated galaxy mass range, and our results may not have fully converged yet. If we perform simulations with larger box sizes ($L>100\,\himpc$), then the $\eta_F$ signal might become even stronger with more massive galaxies in the simulated volume.

\begin{deluxetable*}{lcccccc}[h]
\tablecaption{Simulation Parameters}
\tablehead{
\colhead{Simulations} & \colhead{Box Size} & \colhead{$N_{\rm ptcl}$}&
\colhead{$m_{\rm DM}$} & \colhead{$m_{\rm gas}$} & \colhead{$\epsilon_g$}&
\colhead{$h_{\rm min}$} \\
  & \colhead{[$\hicmpc$]} &  & \colhead{[$\himsun]$} & \colhead{[$\himsun$]} & \colhead{[$\hickpc$]} & \colhead{[$h^{-1}\,{\rm pc}$]}  
}
\startdata
L100N512 (Osaka20) & 100 & $2\times 512^3$ & $5.38\times 10^8$ & $1.00\times 10^8$ & 7.8 & 260 \\
L50N256 & 50 & $2\times 256^3$ & $5.38\times 10^8$ & $1.00\times 10^8$ & 7.8 & 260 \\
\hline
L40N512 & 40 & $2\times 512^3$ & $3.44\times 10^7$ & $6.43\times 10^4$ & 2.6 & 87 \\
\enddata
\tablecomments{Parameters of the simulations used for resolution and box-size test. The L100N512 simulation corresponds to the Osaka20 runs listed in Table~\ref{tab:sim}. The listed parameters are as follows: $N_{\rm ptcl}$ is the total number of particles (dark matter and gas), $m_{\rm DM}$ is the dark matter particle mass, $m_{\rm gas}$ is the initial mass of gas particles (which may change over time due to star formation and feedback), $\epsilon_g$ is the comoving gravitational softening length, and $h_{\rm min}$ is the minimum physical gas smoothing length at $z=2$ (see \S~\ref{sec:sim}). 
}
\label{tab:res}
\end{deluxetable*}

\begin{deluxetable}{lcccc}[h]
\tablecaption{Galaxy Counts}
\tablehead{
\colhead{Simulations} &  & \colhead{$N_{\rm gal}$} &  &  \\
  & \colhead{$10^8$--$10^9$} & 
  \colhead{$10^9$--$10^{10}$} & 
  \colhead{$10^{10}$--$10^{11}$} & 
  \colhead{$>$$10^{11}\Msun$}
}
\startdata
L100N512 (Osaka20) & 8344 & 13893 & 3200 & 148 \\
L50N256 & 1039 & 1705 & 425 & 14 \\
\hline
L40N512 & 9476 & 3036 & 706 & 14 
\enddata
\tablecomments{$N_{\rm gal}$ is the number of galaxies in each galaxy stellar mass range. 
}
\label{tab:galcount}
\end{deluxetable}

% Fig. 13

% Fig. 14

% Fig. 15

%%%%%%%%%%%%%%%%%%%%%%%%%%%%%%%%%%%%%%%%%%%%%%%%%%%%%%

\section{Variation in $\eta_F$ due to Redshift Offset} 
\label{app:gaussian}

\begin{figure}[ht]
    \epsscale{1.1}
    \plotone{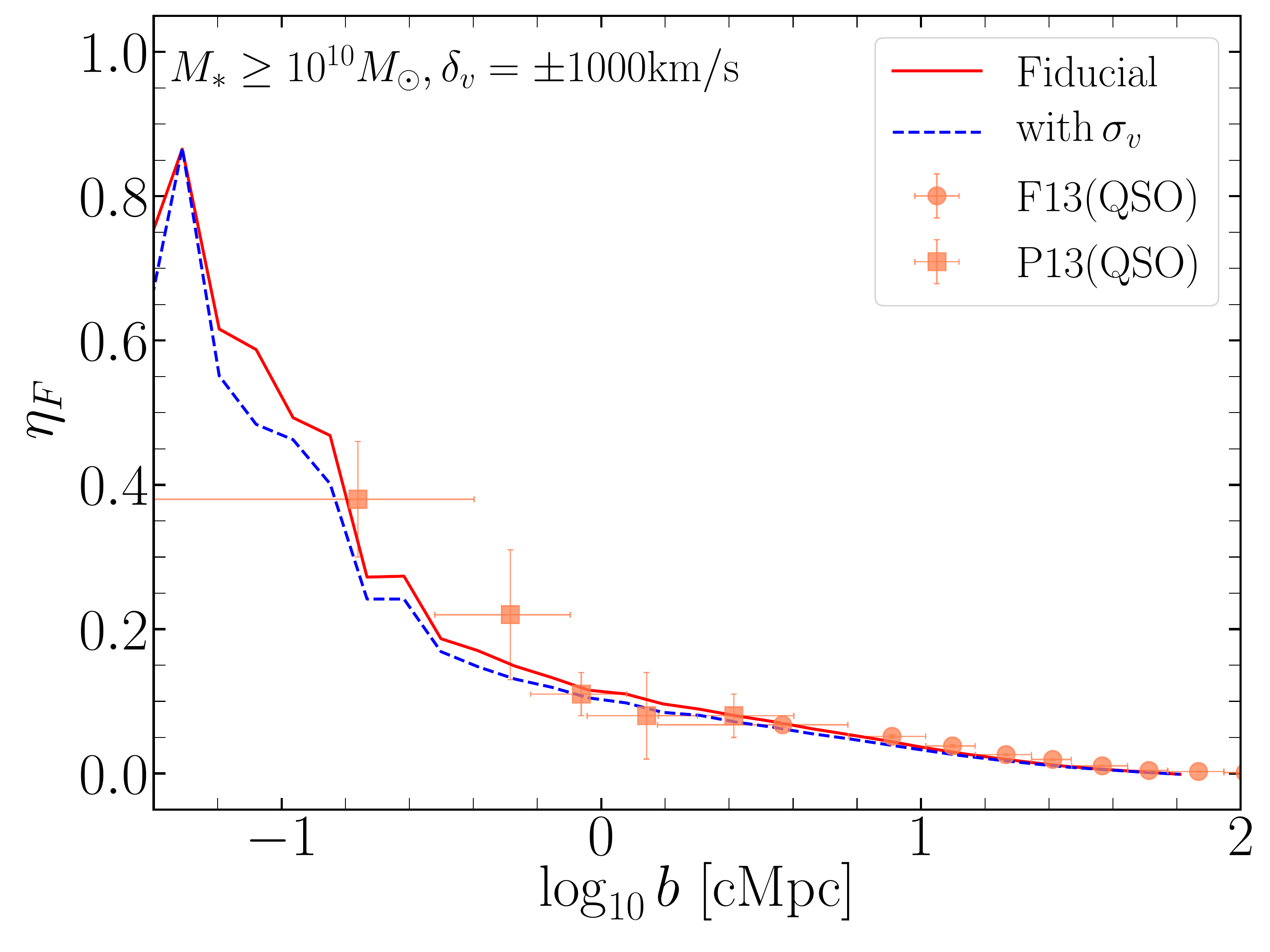}
    \caption{ Flux contrast with and without a Gaussian scatter of $\pm 500$\,km\,s$^{-1}$ along the LOS for the galaxy redshifts (blue dashed and red solid lines, respectively).  The red line shows the same data as the one in Figure~\ref{fig:impact}(c) for $\Mstar > 10^{10}\Msun$. 
    } 
    \label{fig:gaussian}
\end{figure}

Following the previous works \citep{Font-Ribera13,Pro13,Meiksin17}, we examine the impact of redshift errors of galaxies on $\eta_F$ by introducing a Gaussian scatter in the galaxy redshift along the LOS. 
Figure~\ref{fig:gaussian} shows the flux contrast with and without a Gaussian scatter of $1\sigma= \pm 500$\,km\,s$^{-1}$ along the LOS for the galaxy redshifts.  
We find that the effect of redshift offset is not so strong, and only decreases the $\eta_F$ signal by about 10--15\%. This is reasonable given that the velocity range of $\delta_v=\pm 1000$\,km\,s$^{-1}$ is greater than the majority of Gaussian scatter.

% Fig. 16

%%%%%%%%%%%%%%%%%%%%%%%%%%%%%%%%%%%%%%%%%%%%%%%%%%%%%%

\newpage

\bibliographystyle{aasjournal}
\bibliography{master}

%% This command is needed to show the entire author+affilation list when the collaboration and author truncation commands are used.  It has to go at the end of the manuscript.

%% Include this line if you are using the \added, \replaced, \deleted  commands to see a summary list of all changes at the end of the article.

\end{document}